\def\Mpc{\,{\rm Mpc}}
\def\kpc{\,{\rm kpc}}
\def\pc{\,{\rm pc}}
\def\Msun{\,{\rm M_\odot}}
\def\Gyr{\,{\rm Gyr}}
\def\Myr{\,{\rm Myr}}
\def\kms{\,{\rm km}\,{\rm s}^{-1}}

\documentclass[usenatbib]{mn2e}
\usepackage{graphicx}
\usepackage{aas_macros}
\begin{document}

\title[Massive black holes in dwarf spheroidal galaxy haloes?]{Massive black holes in dwarf spheroidal galaxy haloes?}

\author[Jin et al.]{Shoko Jin$^{1,2}$\thanks{e-mail: shoko@ast.cam.ac.uk}, Jeremiah~P.~Ostriker$^1$ \& Mark~I.~Wilkinson$^1$\\
$^1$Institute of Astronomy, University of Cambridge, Madingley Road, Cambridge, CB3 0HA, U.~K.\\
$^2$Clare College, University of Cambridge, Trinity Lane, Cambridge,
CB2 1TL, U.~K.}

\date{Accepted for publication in MNRAS}
\maketitle

\begin{abstract}
It is now established that several of the Local Group dwarf Spheroidal
galaxies (dSphs) have large mass-to-light ratios. We consider the
possibility that the dark matter in the haloes of dSphs is composed of
massive black holes with masses in the range $10^{5}\Msun$ to
$10^7\Msun$. We use direct $N$-body simulations to determine the long
term evolution of a two-component dSph composed of a
pressure-supported stellar population within a black hole dominated
halo. The black holes are initially distributed according to a
Navarro, Frenk \& White profile. For black hole masses between
$10^5\Msun$ and $10^6\Msun$, the dark matter halo evolves towards a
shallower inner profile in less than a Hubble time. This suggests that
black holes in this mass range might provide an explanation for the
origin of the dark matter cores inferred from observations of Low
Surface Brightness galaxy rotation curves. We compare the simulated
evolution of the stellar population with observed data for the Draco
dSph and find that dynamical heating generally leads to the rapid
dispersal of the stellar population to large radii. The dependence of
the heating rate on the black hole mass is determined, and an upper
limit of $10^{5}\Msun$ is placed on the individual black holes
comprising the dark matter halo of Draco, if its present stellar
distribution is representative of the initial one. We also present a
simple scaling argument which demonstrates that the dynamical heating
of an initially compact, though not self-gravitating, stellar
distribution might produce a remnant distribution similar in extent to
Draco after 10 Gyr, even for black hole masses somewhat in excess of
$10^{5}\Msun$.
\end{abstract}
\begin{keywords}dark matter --- galaxies: dwarf, haloes --- methods: $N$-body simulations
\end{keywords}

\section{Introduction}
\label{sec:intro}

The existence of dark matter in the Universe, which has an abundance
six times that of baryonic matter, is inferred from sources as diverse
as the disk rotation curves of spiral galaxies, gravitational lensing
studies of galaxy clusters, the very high mass-to-light ratios of
nearby dwarf galaxies and analysis of Cosmic Microwave Background
fluctuations.  However, the actual composition of the dark matter in
the Universe is as yet unknown ~\citep{Jerry2003}. One method by which
dark matter may be detected and its properties probed is by means of
substructure lensing ~\citep[e.g.][]{Metcalf2001,Li2002,Keeton2003}.
Dark matter studies in the Local Group can also yield constraints on
dark matter models via the mass density as inferred from kinematic
measurements ~\citep[e.g.][]{Mark2002}.

The standard cold dark matter (CDM) cosmology, though popular and
well-tested on large scales, has some unresolved issues.  The two best
known are the over-abundance of low-mass satellite haloes identified at
the present epoch in cosmological simulations relative to the number
of observed Local Group satellites ~\citep[e.g.][]{Moore1999} and the
much-debated inner profile of CDM haloes
~\citep[e.g.][]{Blok2002,Swaters2003}.  The latter may be due to
resolution issues in the simulations, where a range of inner density
slopes are found by various authors~\citep[see][ for a concise
summary]{Suto}. Low mass haloes may also have intrinsically softer
inner profiles than more massive systems
~\citep{Massimo2003}. Observationally, there is considerable evidence
that massive galaxies also do not have cusped haloes, but instead
contain roughly uniform density
cores~\citep[e.g.][]{binev01,salucci01,bsd03}. The issue of the
missing satellite haloes may be due to the fact that some of the
satellite galaxies are very diffuse; a few bound stellar systems may
therefore be undetected simply due to their low surface brightness
~\citep{Mateo1998}.  It has also been suggested that the satellite
crisis can be resolved if all the Local Group dwarf spheroidal
galaxies (dSphs) lie in significantly more massive haloes than those
implied by previous estimates, which were based only on their central
velocity dispersions~\cite[e.g.][]{Stoehr2002,Hayashi2003}. However,
differences between the radial distribution of massive satellite haloes
about their parent galaxies in simulations and the observed
distribution of the Milky Way dSphs suggest that this cannot be the
complete solution to this problem~\citep[e.g.][]{DeLucia03}. Another
possibility is that there may exist dark matter haloes which do not
harbour a stellar population.

In order to address these issues, some alternatives to CDM have been
proposed: these include warm dark matter and self-interacting dark
matter ~\citep[e.g.][]{SS2000,Gnedin_Ostriker}.  An interesting
complete alternative to dark matter has also been suggested for the
observed dwarf Spheroidals (dSphs) whereby special phase-space
characteristics permit the long-lived remnant of a tidally disrupted
satellite to assume the appearance of a dSph in projection
~\citep{Klessen_Kroupa}; however, difficulties in reproducing the data
on Draco have recently been noted in this model ~\citep{Klessen2003}.

There has been considerable interest in cosmological CDM simulations
in the past decade, with a variety of halo profile fits being
presented by different authors.  Following a set of large cosmological
simulations, ~\cite{NFW96,NFW97} proposed a universal density profile
to accommodate virialised CDM haloes of all masses, which has become
commonly known as the NFW profile.  The density in the inner region is
cusped as $r^{-1}$, whereas at larger radii it is proportional to
$r^{-3}$, with the location of the transition being defined by a scale
radius used to characterise the profile.  Several authors have
subsequently confirmed the fit provided by the NFW profile, although
others have also presented steeper inner slopes for the halo density
~\citep[e.g.][]{Fukushige2001,Ghigna2000,Colin2003}.  A generalised
density law has therefore been suggested, for which the original NFW
profile is a special case, in order to account for such
findings~\citep[e.g.][]{Zhao1996}.  As long as the dark matter is cold
and collisionless, the halo profile would not be expected to depend on
the nature of the dark matter and, more specifically, it should be
independent of the dark matter particle mass.  This remains true
unless we consider particles whose masses are much greater than a
solar mass, for which two-body dynamical effects would become
relevant.  Such phenomena would first be observed in the lowest mass,
lowest velocity dispersion systems such as the dSphs.

Much interest in recent years has been focused on the study of the
Local Group satellites, not least because their relative proximity has
led to the increased availability of high quality data
~\citep{IH95,Mateo1998}, which place stricter constraints on the
nature of the dark matter that appears to dominate their gravitating
mass.  These systems have also received much attention on the
modelling and simulation front ~\citep[e.g.][]{Stoehr2002,Mark2002}.
Amongst those with the highest mass-to-light ratios is the Draco dSph
galaxy.  At a heliocentric distance of approximately $80\kpc$, Draco has
been studied extensively and a wealth of kinematic data exists; in
particular, measurements of stellar velocities out to large radii are
now available ~\citep[]{Jan2002,Mark2004}.  Given that within the
region probed by the stellar distribution it also appears to have
suffered relatively little from the tidal effects of the Milky Way
~\citep{Odenkirchen,Klessen2003,Mark2004}, unlike some of the other
Local Group satellites such as the Sagittarius dwarf
~\citep[e.g.][]{Helmi2001}, Draco is an ideal candidate with which to
study the possible properties of dark matter.  Draco is also known to
contain an old stellar population with an age of more than 10 Gyr; any
model of this galaxy must therefore remain stable for at least this
length of time in order to be compatible with this observation.

One conceivable candidate for the dark matter, either as the sole or
major component, is an ensemble of massive black holes.  These could
be primeval, or may have grown by accretion and mergers from a
primeval population, or have origins in later non-linear structures.
~\citeauthor{LO85} (1985: hereafter LO85) investigated the way in
which a halo of massive black holes would affect the stellar dynamics
in the Milky Way and obtained results that are generally consistent
with observations.  Our present study investigates the dynamical
effects of a dark matter halo composed purely of primordial massive
black holes on the evolution of the stellar component of a dwarf
galaxy.  It was shown in LO85 that for a dSph such as Draco, an upper
limit for the individual black hole mass would be $2\times10^6\Msun$,
based upon stability arguments for both the stellar population and the
dark matter halo forming the galaxy.  More recently, \cite{mao04}
have suggested that massive black holes with masses in the range
$10^5-10^6\Msun$ may offer a viable explanation for the observed image
flux ratios in strong gravitational lens systems. We use direct
$N$-body simulations to trace the evolution of a model dSph whose
initial structural parameters are chosen such that they are compatible
with present-day observations of Draco.  The stellar population is
treated as a cluster of tracer particles, whose evolution is
dependent on the potential of the dark halo.  The distribution of the
halo particles is chosen to follow an NFW profile, whilst the stellar
population is given an initial distribution which is described by a
Plummer law.  We assume the dark matter of the halo to consist
entirely of massive black holes, formed prior to the formation of the
stellar population. Their masses are chosen to cover a range whose
maximum value has, for generality, been taken to be larger than the
constraints imposed by LO85; in each simulation, all black holes have
the same mass.

In Section~\ref{sec:sim} we discuss the generation of our simulated
models.  We present the results in Section~\ref{sec:results} along
with a discussion in Section~\ref{sec:disc}.  In
Section~\ref{sec:init_cond} we discuss alternative initial conditions
for the stellar distribution of dSphs and their implications for black
hole haloes. Finally we present a summary and our main conclusions in
Section~\ref{sec:con}.

\section{Simulations}
\label{sec:sim}

\begin{table*}
\caption{
Model parameters:\label{tab:sim}
Column 1 gives the simulation number; columns 2--4 give the total
number of black holes, the mass of a single black hole and the total
halo mass respectively; column 5 gives the total number of 
tracers; column 6 gives an estimate of the initial half-mass
relaxation time $t_{\rm rh}$ given by equation~(\ref{eq:t_rh}); column
7 gives the corresponding relaxation time for the central 5 percent
(by mass) of the black hole distribution; columns 8--10 give the halo
scale length, density scale and concentration respectively; column 11
gives the length conversion from $N$-body units to physical distances
in kpc; column 12 gives the gravitational softening $\epsilon$ (in
$N$-body units) used.}
\begin{tabular}{lrrccrrccccl}
\hline
&              
& \multicolumn{1}{c}{$M_{\rm BH}$} 
& \multicolumn{1}{c}{$M_{\rm total}$} 
&           
& \multicolumn{1}{c}{$t_{\rm rh}$} 
& \multicolumn{1}{c}{$t_{\rm r0}$} 
& \multicolumn{1}{c}{$r_{\rm s}$}  
& \multicolumn{1}{c}{$\rho_{\rm s}$}      
&            
& \multicolumn{1}{c}{$r_{\rm phys}$} 
& \\
 Model   
& \multicolumn{1}{c}{$N_{\rm BH}$} 
& \multicolumn{1}{c}{($10^6\Msun$)}
& \multicolumn{1}{c}{($10^9\Msun$)} 
& \multicolumn{1}{c}{$N_{\rm tr}$} 
& \multicolumn{1}{c}{(${\rm Gyr}$)}
& \multicolumn{1}{c}{(${\rm Gyr}$)} 
& \multicolumn{1}{c}{(kpc)} 
& \multicolumn{1}{c}{($10^5\rho_{\rm crit}$)} 
& \multicolumn{1}{c}{$c$}       
& \multicolumn{1}{c}{(${\rm kpc}$)} 
& \multicolumn{1}{c}{$\epsilon$} \\ 
\hline \hline
1...... & 1001 & 1.0 & 1.0 & -- & 3.4 & 0.07 & 0.6 & 10.4 & 32.6 & 5.5 & 0.02 \\
2...... & 101 & 10.0 & 1.0 & $10^4$ & 0.5 & 0.02 & 0.6 & 10.4 & 32.6 & 4.3 & 0.02 \\
3...... & 300 & 3.3 & 1.0 & $10^4$ & 1.2 & 0.05 & 0.6 &  10.4 & 32.6 & 4.9 & 0.02 \\
4...... & 1001 & 1.0 & 1.0 & $10^4$ & 3.4 & 0.07 & 0.6 & 10.4 & 32.6 & 5.5 & 0.02 \\
5...... & 1001 & 1.0 & 1.0 & $10^5$ & 3.4 & 0.08 & 0.6 & 10.4 & 32.6 & 5.5 & 0.0002 \\
6...... & 1001 & 1.0 & 1.0 & $10^5$ & 3.4 & 0.08 & 0.6 & 10.4 & 32.6 & 5.5 & 0.02 \\
7...... & 3001 & 1.0 & 3.0 & $10^5$ & 7.3 & 0.19 & 1.2 & 5.2 & 16.3 & 7.1 & 0.02 \\
8...... & 3000 & 0.3 & 1.0 & $10^4$ & 8.5 & 0.12 & 0.6 & 10.4 & 32.6 & 5.1 & 0.02 \\
9...... & 3000 & 0.3 & 1.0 & $10^5$ & 8.5 & 0.12 & 0.6 & 10.4 & 32.6 & 5.1 & 0.02 \\
10.... & 9002 & 0.3 & 3.0 & $10^5$ & 19.5 & 0.45 & 1.2 & 5.2 & 16.3 & 7.2 & 0.02 \\
11.... & 10001 & 0.1 & 1.0 & $10^4$ & 24.9 & 0.49 & 0.6 & 10.4 & 32.6 & 5.2 & 0.02\\
\hline
\end{tabular}
\end{table*}

A summary of the simulations performed is given in
Table~\ref{tab:sim}. The simulations were carried out using the NBODY2
code ~\citep{Sverre}, modified to incorporate tracer particles
orbiting within the underlying  potential of the dark matter halo to
represent the observed stars.  The mass of the system was assumed to
be derived entirely from the dark matter.  The black hole mass was
varied in different runs over the range $10^{5}$--$10^{7}{\rm
M_{\odot}}$.  We adopt the standardised $N$-body units described by
~\cite{Heggie}: $G=1$, $M_{\rm tot}=1$, where $M_{\rm tot}$ is the
total self-gravitating mass, and the initial total energy
$E_{0}=-1/4$.  The inclusion of a gravitational softening parameter
$\epsilon$, or softening length, ensures that force singularities do
not occur as the separation of particles tends to zero.  A value of
$\epsilon = 0.02$ in $N$-body units, chosen for computational
expediency, was used for the majority of our simulations,
corresponding to a physical length of $86\pc<\epsilon<144\pc$
depending on the model.  Model 5 was performed as a repeat of model 6,
but with the softening length reduced by a factor of 100, in order to
verify that we had not missed any significant physical effects as a
result of using relatively large softening in the other models.  The
number of tracers at the start of the simulation was 10,000 for models
2--4, 8 and 11, and 100,000 for the remaining models. All simulations
were performed on the Sun Workstation Cluster at the Institute of
Astronomy and allowed to run to at least 20 Gyr.

\subsection{Initial Conditions}
\begin{figure}
  \resizebox{0.5\textwidth}{!}{
  \includegraphics[]{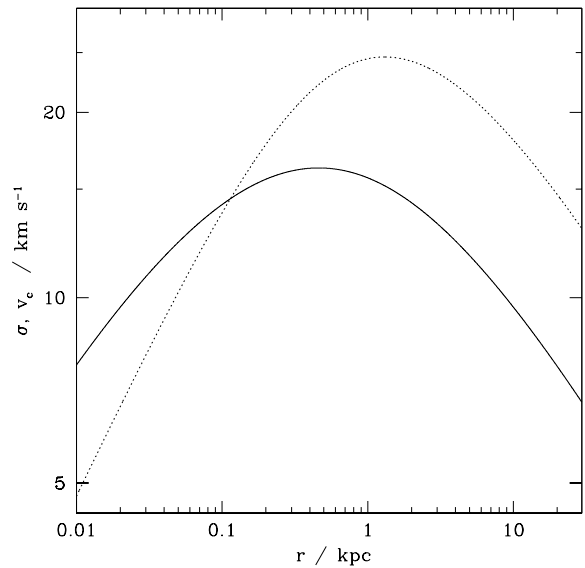}} \caption{The velocity
  dispersion $\sigma$ (solid curve) and circular velocity $v_{\rm c}$
  (dotted curve) of the halo particles as a function of radius. The
  initial velocities of the massive particles are generated from this
  distribution for all models except models 7 and 10 in which the halo
  scale length $r_{\rm s}$ is larger.\label{fig:vdisp}}
\end{figure}

The NFW density profile ~\citep{NFW96,NFW97}, proposed as a universal
fitting formula for CDM haloes which form in a hierarchical clustering
scenario, has the form:

\newcommand{\ud}{\mathrm{d}}
\begin{equation}
  \label{NFW}
  \rho(r) = \frac{\rho_{s}r_{s}^{3}}{r(r+r_{s})^{2}}\,,
\end{equation}
where the parameters $\rho_s$ and $r_s$ used to characterise the NFW
profile are the density scale and scale radius respectively.

The black holes in our simulations are initially distributed according
to the NFW profile. The allocation of the initial particle velocities
is performed by numerically evaluating the velocity dispersion using
the Jeans equation for a spherically symmetric system
~\citep[e.g.][]{BT87},

\begin{equation}
\label{eq:Jeans}
  \frac{1}{\nu}\frac{\ud(\nu \sigma^2)}{\ud r} = -\frac{\ud\Phi}{\ud
  r},\\
\end{equation}
where $\nu(r)$ is the halo density profile, $\sigma(r)$ is the one
dimensional velocity dispersion for the halo and the underlying halo
potential is $\Phi(r)$.  We have assumed velocity isotropy for the
halo particles, which gives a satisfactory fit to the results observed
in cosmological simulations. In this case, $\sigma(r)$ has the form
\begin{equation}
\label{eq:Jeans_sigma}
  \sigma^2=4\pi\,G\,\rho_{\rm s}\,r_{\rm s}^2\,(s_{\rm max}-s_x)\,x\,(1+x)^2,
\end{equation} 
where $x$ is the scaled radius $r/r_s$ and $s_x$ is given by\\

\begin{equation}
s_x =
\int_{0}^{x}\frac{\ln\,(1+x')+(1+x')^{-1}-1}{x'^3\,(1+x')^2}\,\ud
x'\,.
\end{equation}
The value of $s_{\rm max}$ was evaluated by numerically integrating up
to a sufficiently large radius.  The form of the one dimensional dispersion so
derived is shown in Fig.~\ref{fig:vdisp}, which also shows the
circular velocity profile for the NFW halo.

For each particle, we randomly generate a velocity from a Maxwellian
velocity distribution with velocity dispersion calculated using
equation (\ref{eq:Jeans_sigma}) at the position of the particle. While
at each place the total energy (gravitational and kinetic) of the
local particles is correct and the Jeans equation is satisfied, we
note that recent work by~\cite{kmm04} has found that the assumption of
a Maxwellian velocity distribution can lead to a re-adjustment and
short-term spurious evolution in $N$-body simulations, due to the fact
that the true velocity distribution is not an exact
Maxwellian. However, this evolution does not have a significant impact
on the simulations we present in this paper. The initial re-adjustment
takes place on a short time scale (the halo crossing time) while the
effects we are studying take place on the much longer halo relaxation
time scale.

We note in passing that it is possible to relate the maxima of the
rotation curve $v_{\rm max}$ and velocity dispersion $\sigma_{\rm
max}$ of an NFW halo, and the corresponding radii $r_1$ and $r_2$, to
$\rho_{\rm s}$ and $r_{\rm s}$ via
\begin{equation}
  v_{\rm max}^2 = 2.717 G \rho_{\rm s} r_{\rm s}^2\qquad 
\sigma_{\rm max}^2 = 1.185 G \rho_{\rm s} r_{\rm s}^2
\end{equation}
and
\begin{equation}
  r_1 = 2.163 r_{\rm s}\qquad r_2 = 0.763 r_{\rm s}\,.
\end{equation}
The numerical constants were obtained by direct evaluation of the
circular speed and velocity dispersion profiles. As
Fig.~\ref{fig:vdisp} also shows, the maxima of the two profiles do not
occur at the same radius.

We represent the dSph stellar population by tracer particles.
This is justified on the basis that several of the dwarf satellites of
the Milky Way are observed to have very high mass-to-light ratios; in
particular, Draco is known to have an extremely large mass-to-light
ratio of 350--1000 in solar units ~\citep{Jan2001}.  These systems
therefore appear to be very dark matter dominated, with their stellar
populations residing in massive dark haloes.  In our simulations, the
tracers are initially distributed according to a Plummer profile,
which is known to provide a good fit to the present light distribution
in the inner regions of certain dSphs, including Draco.  The form of
the three dimensional~\cite{Plummer1911} luminosity distribution
$\nu(r)$ is given by
\begin{equation}
\nu(r) = \frac{\nu_0 r_{\rm P}^5}{\left(r_{\rm P}^2+r^2 \right)^{5/2}},
\end{equation}
where the scale length $r_{\rm P}$ corresponds to the projected
half-light radius of the model. For Draco, $r_{\rm P}$ is
approximately $232\pc$, which implies a core radius (the radius at
which the surface brightness falls to half its central value) of
$150\pc$. The tracer velocities are again obtained by means of the
Jeans equation~(\ref{eq:Jeans}).  As before, we assume velocity
isotropy, as suggested by the observed stellar kinematics of Draco
~\citep{Jan2002}.

The halo parameters for the majority of our simulations are chosen to
reflect the observational estimates of the mass of
Draco~\citep{Jan2001}. The halo mass is constrained by requiring that
the value of $M_{200}$, the mass contained within a radius $r_{\rm
200}$ which encompasses a mean overdensity of 200 times the critical
density, yields a mass within $3 r_{\rm P}$ which is consistent with
that of Draco. The scale length of the halo is given by $r_{\rm s} =
r_{\rm 200}/c$, where the `concentration' $c$ of the halo is set using
the scaling relation of ~\cite{Bullock2001}:

\begin{equation}
  \label{eq:init1}
  c = 9\left(\frac{M_{\rm 200}}{M_{\ast}}\right)^{-0.13},
\end{equation}
where $M_{\ast}=2.14\times10^{13}\Msun$ is the typical collapsing mass
at the present epoch.  The value of $r_{\rm 200}$ is then given by

\begin{equation}
  r_{\rm 200} = \left(\frac{3M_{\rm 200}}{800\pi\rho_{\rm
  crit}}\right)^{1/3}\,,
\end{equation}
where $\rho_{\rm crit}=3H^2/8\pi G$ is the critical density for
closure of the Universe; we take $H=70 \kms \Mpc^{-1}$.  The relation
between the concentration and the other parameters is made complete by
the alternative description of the density scale as $\rho_{\rm s} =
\delta_c \,\rho_{\rm crit}$ where\\

\begin{equation}
  \delta_{c} = \frac{200}{3}\frac{c^3}{\left[\rm ln (1+c)-c/(1+c)\right]}\,.
\end{equation}
The specification of the halo mass $M_{200}$ thus allows all other
parameters to be deduced using the above relations. While most of our
simulated haloes have similar masses to that of Draco, we also include
a number of haloes with larger scale radii. This extends our modelling
to include Low Surface Brightness (LSB) galaxies.

\subsection{Relaxation Time Scale}
\label{subsec:relax}
We expect the time scale on which the halo profile evolves to be
related to $t_{\rm rh}$, the relaxation time at the half-mass radius
of the initial black hole distribution.  The relaxation time scale is
essentially the time scale over which particles lose memory of their
initial energies due to mutual encounters.  The half-mass relaxation
time for each model, listed in Table
\ref{tab:sim}, is estimated using the definition of $t_{\rm rh}$ as
given by~\citeauthor{BT87} (\citeyear[ chapter~8]{BT87}):

\begin{equation}
\label{eq:t_rh}
t_{\rm rh} = \frac{2.1\times10^9{\rm yr}}{{\rm
ln}(0.4N)}\left(\frac{10^6\Msun}{m}\right)\left(\frac{M}{10^9\Msun}\right)^\frac{1}{2}\left(\frac{r_{\rm
h}}{1\kpc}\right)^\frac{3}{2}\,,
\end{equation}
where $M$ and $N$ are the total mass and number of black holes in the
system respectively, $m$ is the mass of an individual black hole and
$r_{\rm h}$ is the half-mass radius.  We note that the derivation of
equation~(\ref{eq:t_rh}) assumes a Plummer model and is not precisely
valid for an NFW profile for which the ratio of the half-mass radius
to the gravitational radius is about $15$ percent larger than in the
Plummer case. Thus, this corresponds to a deviation in $t_{\rm rh}$ of
less than $5\%$ for our models. However, equation~(\ref{eq:t_rh})
provides a good indication of the typical relaxation time and we have
chosen therefore to keep it in its more familiar form.

The dependence of relaxation time on black hole mass demonstrates the
conservative nature of our choice of initial parameters for the
simulated haloes. Given that massive black holes are, by definition,
manifestations of Cold Dark Matter, one might expect their generic
density distribution to be similar to an NFW profile. However, this is
not necessarily the case. For example, if the dark matter consists
entirely of $10^5\Msun$ black holes, then a primordial halo of total
mass $10^7\Msun$ contains only 100 particles. As we will demonstrate
in the next section, a black hole halo with an NFW density profile
develops a central core via two-body processes on a time scale of
$t_{\rm rh}$. \cite{ostgne96} estimate that $10^7\Msun$ haloes become
non-linear at a redshift of about $z=13$ (see their Figure~1), at
which time their virial radius can be shown to be about
$0.3\kpc$. According to equation~(\ref{eq:t_rh}), the relaxation time
for such a halo $t_{\rm r,7}$ is then about $100\Myr$. If this time
scale is much less than the difference between the times at which
haloes of $10^7\Msun$ and $10^9\Msun$ become non-linear, then we would
expect $10^9\Msun$ haloes to be formed from cored sub-haloes. In this
case the resulting $10^9\Msun$ haloes would have density profiles which
are significantly less cusped than NFW, which would agree with
observations of LSB galaxies. In fact, the $10^9\Msun$ haloes become
non-linear approximately $t_{\rm r,7}$ later than $10^7\Msun$ haloes
and so it is less clear what the form of the resulting density profile
should be. However, the most unfavourable starting point is the NFW
profile and our simulations therefore constitute the strictest test of
the ability of black hole dark matter to produce cored galaxy haloes.

\section{Results}
\label{sec:results}

\subsection{Halo Evolution}
\begin{figure}
  \resizebox{0.5\textwidth}{!}{
\includegraphics[]{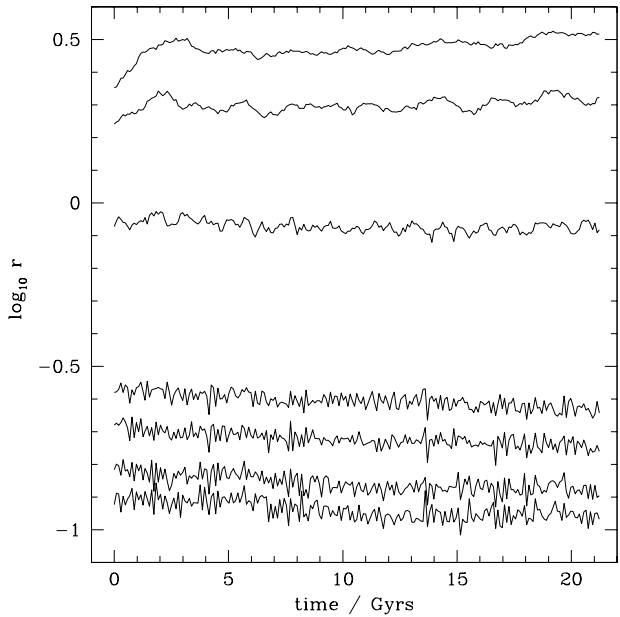}}
\caption{The structural evolution of the halo in model 1 as depicted by the
  Lagrange radii in $N$-body units.  From the lowest curve upwards,
  the radii shown correspond to the following halo mass percentiles:
  7.5, 10, 15, 20, 50, 75 and 85 percent.  The cluster is seen to
  retain its initial structure over the course of the simulation.
 \label{fig:lagr12J}}
\end{figure}

\begin{table}
\caption{Halo parameters at $T=0$ and $T=15\Gyr$:
Column 1 gives the simulation number; column 2 gives the
logarithmic slope of the inner halo at $T=0$ for
those models in which there are sufficient numbers of black holes to
give a statistically meaningful estimate; column 3 gives the radius
$r_{\max}$ at
which the initial circular velocity profile achieves its maximum
value; column 4 gives the maximum value of the initial circular
velocity $v_{\rm c,max}$; columns 5--7: as for columns 2--4 but for the
haloes at $T=15$ Gyr.}\label{tab:summary}
\begin{tabular}{lcccccc}
\hline
 & \multicolumn{3}{c}{$T=0$} &\multicolumn{3}{c}{$T=15$} \\
Model & $\alpha$ & $r_{\rm max}$ & $v_{\rm c,max}$ & $\alpha$ & $r_{\rm max}$ & $v_{\rm c,max}$\\ 
&  & (kpc) & (km\,s$^{-1}$) &  & (kpc) & (km\,s$^{-1}$)\\ 
\hline\hline
1......  & - & 1.5 & 24.3 & - & 1.5 & 24.2\\
2......  & - & 1.3 & 25.5 & - & 0.1 & 48.4\\
3......  & - & 1.7 & 25.1 & - & 0.8 & 27.6\\
4......  & - & 1.6 & 24.3 & - & 1.7 & 24.6\\
5......  & - & 1.6 & 24.2 & - & 1.0 & 26.7\\
6......  & - & 1.6 & 24.2 & - & 1.7 & 25.3\\
7......  & - & 2.4 & 30.2 & - & 2.4 & 35.5\\
8......  & 1.33 & 1.3 & 25.1 & 0.22 & 1.3 & 26.0\\
9......  & 1.15 & 1.2 & 25.1 & 0.00 & 1.2 & 25.9\\
10....  & 0.15 & 2.6 & 34.7 & 0.00 & 2.3 & 35.9\\
11....  & 1.12 & 1.2 & 24.6 & 0.33 & 1.3 & 25.6\\
\hline
\end{tabular}
\end{table}

The stability of a cluster of massive black holes with an NFW profile
(and no tracers) was initially assessed in order to determine the
validity of our assumption that such a cluster could act as a halo
enveloping a stellar population (model~1).  As expected from the
estimate of $t_{\rm rh}$ in Table \ref{tab:sim}, we find that the
cluster retains its initial structure over the duration of the
simulation (Fig.~\ref{fig:lagr12J}).  The inner regions are not
reproduced in this plot: their evolution will be discussed in detail
below.

\begin{figure}
  \resizebox{0.5\textwidth}{!}{
\includegraphics[]{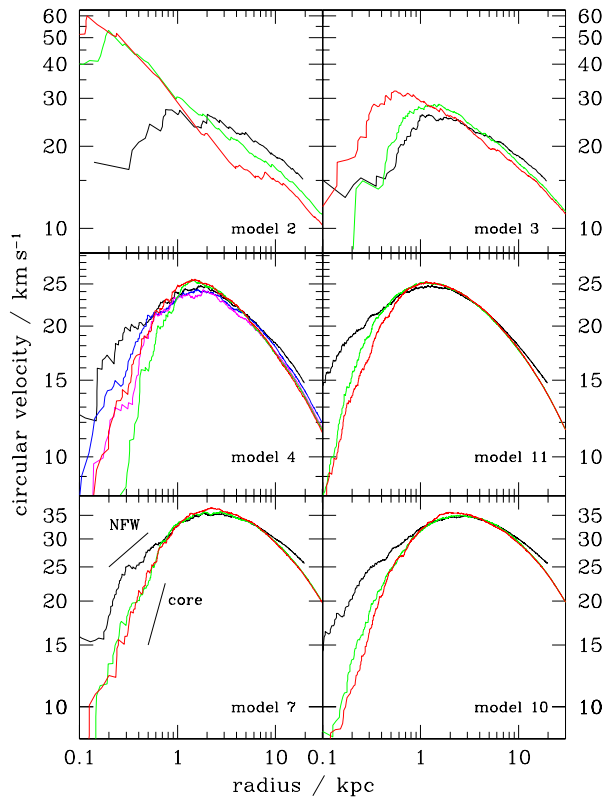}}
\caption{Halo rotation curves for models 2--4, 7, 10 and 11 shown at times $T =
0$ (black), 10 (green) and 20 (red) Gyr.  The plot for model 4 also
shows the circular velocities at 1 (blue) and 2 (magenta) Gyr.  In the
panel for model~7, the labelled lines indicate the expected slope for
an NFW and a cored density profile. \label{fig:vcirc}}
\end{figure}

\begin{figure}
  \resizebox{0.5\textwidth}{!}{
\includegraphics[]{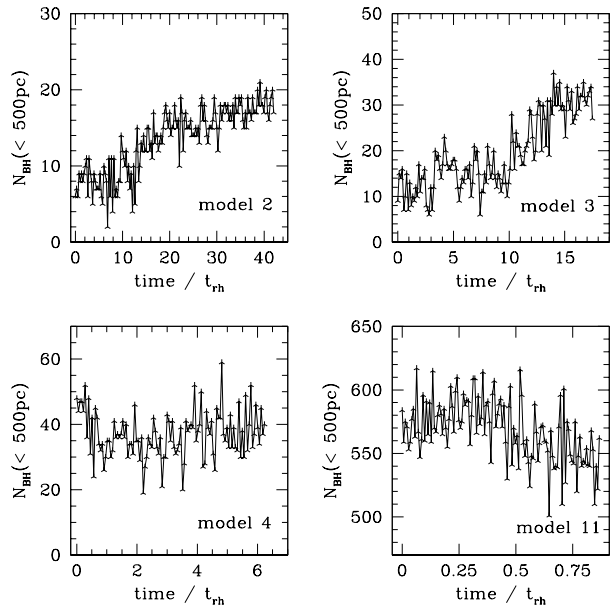}}
\caption{The evolution of the number of black holes within $500\pc$ of the halo
centre for models 2, 3, 4 and 11. The time axes of the plots have been
scaled individually according to the half-mass relaxation time for
each model.  \label{fig:M<500pc}}
\end{figure}

The evolution of the circular velocity profiles for models 2--4, 7, 10
and 11 is shown in Fig.~\ref{fig:vcirc}.  Table \ref{tab:summary}
summarises the properties of these curves.  As will be discussed in
Section~\ref{sec:halo_disc}, the initial evolution leads to the
destruction of the NFW cusp and the formation of a shallower
core. This is followed by slow evolution towards core collapse -- the
time scale for this evolution varies according to the mass of the
black holes. 
The inner slope of the velocity profile becomes steeper for models 7,
10 and 11, indicating a transition from the initial cusp of the NFW
profile to a more uniform density core in the central region of the
halo.  The evolution of the mass within $500\pc$ of the halo centre as
a function of the initial half-mass relaxation time is shown in
Fig.~\ref{fig:M<500pc} for models 2--4 and 11. Models 2 and 3 display
a steady increase in the mean density within $500\pc$ while in model
4, which has better time resolution (at early times) due to its longer
$t_{\rm rh}$, the density first decreases on a time scale of $t_{\rm
  rh}$. This initial decrease in the mass within $500\pc$ is followed
by a slow increase as the system evolves towards core collapse. In
model 11, for which $20\Gyr$ corresponds to less than one relaxation
time, the central density is still decreasing at the end of the
simulation. \cite{Hayashi2003} observe similar evolution in their
simulations of CDM haloes. They find that this variation of the
central density profile occurs on the local escape time scale which is
approximately $136$ times the local (i.e. central) relaxation time
scale. As we will discuss later, we find that the evolution proceeds
with a characteristic time scale comparable to $t_{\rm rh}$, which is
similar in magnitude to the evolutionary time scale observed
by~\cite{Hayashi2003}.

Fig.~\ref{fig:vcirc} shows that, after $20\Gyr$, the evolution towards
core collapse is well advanced only for models 2 and 3. At the
opposite end of the mass scale, model 11 has only just developed a
core by about $10\Gyr$. For model 4, we note that although the inner
profile begins to steepen again after $10\Gyr$ of evolution, the
circular speed curve after $20\Gyr$ indicates that the inner regions
remain less cusped than the initial conditions. We conclude that core
collapse in black hole haloes will not occur within a Hubble time
unless the black hole mass is greater than $10^{6.5} \Msun$. Most of
the models with black hole masses below this value have the attractive
feature that the slope of the cusp
\textit{decreases} with time, with the final approximate inner slope
being shown in Table
\ref{tab:summary} for those models for which a statistically
meaningful estimate can be made. Thus, NFW haloes composed of black
holes with masses in the range $10^5\Msun$ to $10^{6.5} \Msun$ can
develop central cores in less than a Hubble time. This may provide
an explanation for the origin of the apparent cores in the density
profiles of LSB galaxies.

\subsection{Evolution of the Stellar Population}
\begin{figure}
  \resizebox{0.5\textwidth}{!}{
\includegraphics[]{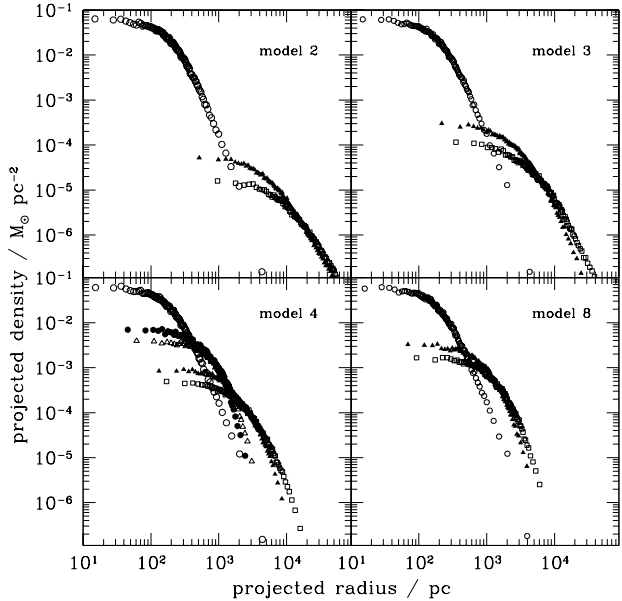}}
\caption{Projected tracer density plots for models 2, 3, 4 and 8 shown at times
$T =0$ (open circles), 10 (filled triangles) and 20 (open squares)
Gyr.  The plot for model 4 also shows the densities at sampling times
of 1 and 2 Gyr with filled circles and open triangles
respectively. \label{fig:projdens}}
\end{figure}

\begin{figure}
  \resizebox{0.5\textwidth}{!}{
\includegraphics[]{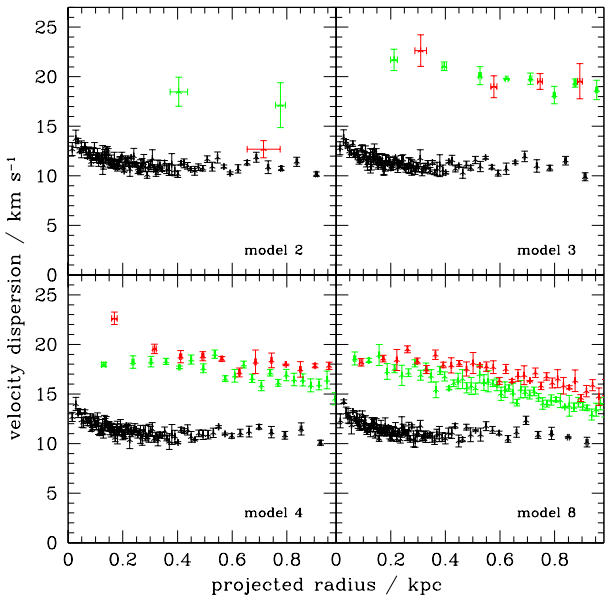}}
\caption{Line-of-sight velocity dispersion plots for models 2, 3, 4 and 8 shown
at times $T =0$ (black), 10 (green) and 20 (red) Gyr.
\label{fig:losvel}}
\end{figure}

\begin{figure}
  \resizebox{0.5\textwidth}{!}{
\includegraphics[]{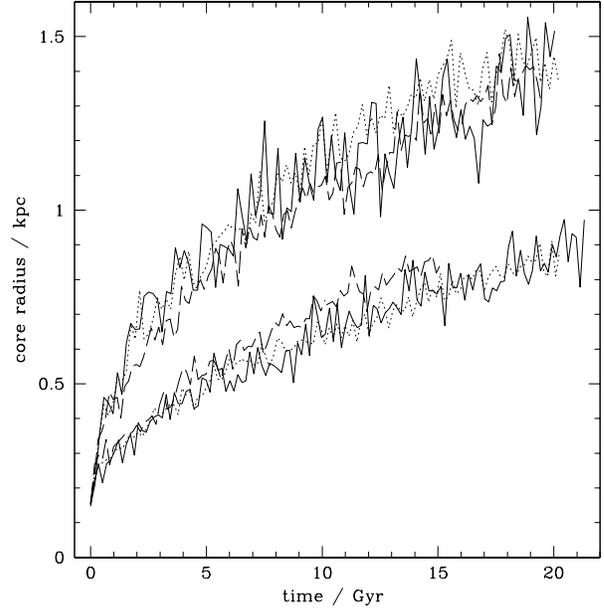}}
\caption{The tracer core radius shown as function of time.  The upper curves
  show models 4 (solid), 6 (dotted) and 7 (dashed); the lower curves
  refer to models 8 (solid), 9 (dotted) and 10 (dashed).  Note that
  the observed core radius of Draco is $\sim$$150
  \pc$.\label{fig:light}}
\end{figure}

\begin{figure}
  \resizebox{0.5\textwidth}{!}{
\includegraphics[]{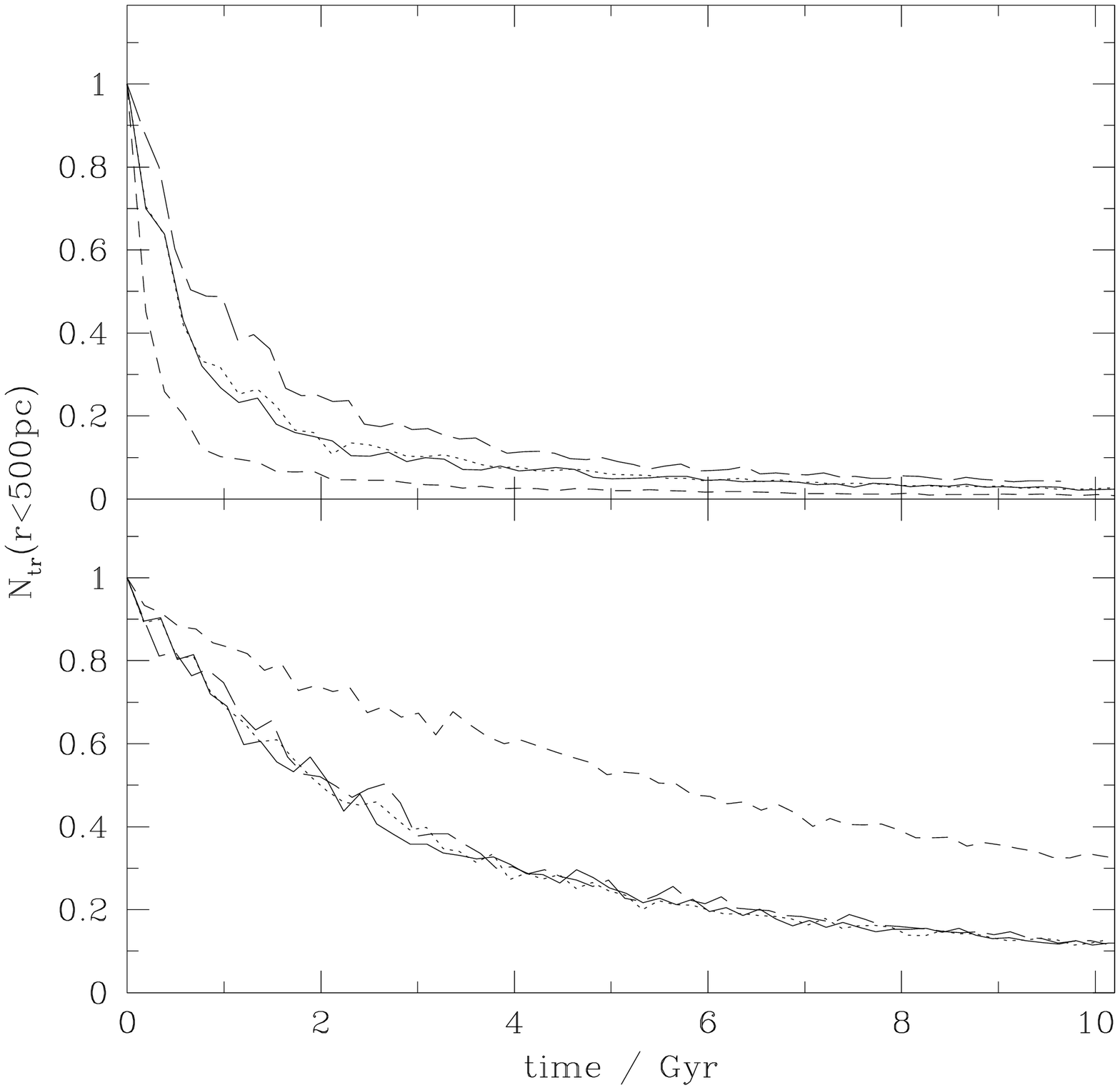}}
\caption{Plots showing the decline in tracer number within $500\pc$.  Top
panel: model 4 (solid), model 5 (short dashed), model 6 (dotted),
model 7 (long dashed). Bottom panel: model 8 (solid), model 9
(dotted), model 10 (long dashed) and model 11 (short dashed).  All
models shown in the top panel have black holes of mass $10^6\Msun$;
those depicted in the bottom panel have black holes of mass
$10^{5.0-5.5}\Msun$.  Model 4 has $10^4$ tracers and black holes of
mass $10^6\Msun$; model 6 is a repeat of model 4 with $10^5$ tracers;
model 5 is a repeat of model 6 with the softening reduced by a factor
of 100; model 7 is a repeat of model 6 where the halo scale length has
been doubled.  Model 8 has $10^4$ tracers and black holes of mass
$10^{5.5}\Msun$; model 9 is a repeat of model 8 with $10^5$ tracers;
model 10 is a repeat of model 8 where the halo scale length has been
doubled; in model~11 the black holes have mass $10^{5}\Msun$.
\label{fig:trcount}}
\end{figure}

The evolution of the tracer density profile for some of our models is
shown in Fig.~\ref{fig:projdens} where we have assumed each star to
have a physical mass of 1$\Msun$.  In evaluating the projected
densities, we have taken radial bins containing 100 tracer particles
in each of the three projected (Cartesian) planes.  An average value
was then calculated for each bin over all three planes.  We adopt a
similar approach in evaluating the line-of-sight velocity dispersions
shown in Fig.~\ref{fig:losvel}.  The tracer profile becomes extended
over the course of the simulation as the particles acquire energy --
the heating process is apparent from Fig.~\ref{fig:losvel} where the
dispersion is clearly greater for later times.  From the evolution of
the density in projection, it is straightforward to derive the
behaviour of the core radius as a function of time; this is shown in
Fig.~\ref{fig:light} for models 4 and 6--10.  As Fig.~\ref{fig:light}
shows, in all cases the core radius rapidly increases to values which
are inconsistent with the observed value of $150\pc$ for
Draco. Complementary to this is Fig.~\ref{fig:trcount} which shows the
decline of the normalized tracer count within $500\pc$ as a function
of time for several models.  Unsurprisingly, reduced softening is seen
to increase the rate at which the stars are heated; the same effect is
seen if we increase the black hole mass.  For models utilising
$10^6\Msun$ black holes, an increase in the halo scale length reduces
the rate of heating, whereas for smaller mass black holes, the effect
is negligible.  As expected, simply increasing the size of the stellar
population by a factor of 10 produces no effect as regards the tracer
evolution, except that we obtain a better statistical description of
the profile.  In Section~\ref{subsec:draco} we will discuss how a
comparison of the core radii of the simulated tracer populations with
that of Draco can be used to place a constraint on the mass of
individual black holes in the halo of Draco.

The majority of our simulated systems produced no tracer--black hole
binaries, although we observe steady black hole--black hole binary
formation at all times for all of our models.  Tracer--black hole
binaries were, however, observed in models 6, 7, 9 and 10, where the
first two models produced two binaries each during the course of the
simulation, whilst in the latter two models, six and thirteen
tracer--black hole binaries were observed respectively.  Of these,
both models showed one long-lived binary each, corresponding to a
bound system lasting up to 340 and 330 Myr respectively.  Binaries
composed of two black holes were observed frequently and consistently
in all simulations, with typically 2--4 transient binaries per
sampling time, which corresponds to a stable bound system lasting for
approximately 170 Myr (depending on the model); occasionally a binary
remained intact for several Gyr.  Simply reducing the softening (but
not down to zero) did not appear to increase the likelihood of
tracer--black hole binary formation, as exemplified by model 5 which
showed no tracer bound to a black hole at any time during the course
of the simulation.  This may be due to two counteracting effects; a
reduced softening will allow for closer encounters, but at the same
time will increase the heating of the tracer population.  This reduces
the central density of tracers more rapidly than for a system with a
larger softening and decreases the probability of the close encounters
which result in the formation of a bound system.

Given that our simulations were performed using non-zero force
softening, it follows that the rate of binary formation and the
dynamical significance of those binaries that do form cannot be
properly determined from these models.  In order to address these
issues, we have re-simulated models 4 and 11 using the NBODY4
code~\citep{Aarseth:99} on the GRAPE-6 special-purpose computer board
\citep{Makino:97} at the Institute of Astronomy, Cambridge.  The
results are discussed in Section~\ref{subsec:grape} below.

We note that in our simulations using the NBODY2 code we treat the
stars as tracer particles within the potential of the black hole
cluster, thereby allowing improved efficiency in integration time.  It
then follows that the black holes do not `see' the stars; they simply
feel the potential due to the other black holes that are present in
the cluster.  The stars are tracers within this dark matter potential
and their dynamics are unaffected by the stellar population, hence we
do not expect to observe tracer-tracer binaries.  Note also that all
particles are point masses with no internal structure, and hence the
formation of a bound two-body system requires a three-body
interaction, with the third particle carrying away the excess
momentum.

\section{Discussion}
\label{sec:disc}
\subsection{Halo Evolution}
\label{sec:halo_disc}

In this section, we use simple dynamical models to explain the
evolutionary trends of both the black holes and tracers in our
simulations. There is an initial decrease in the number of massive
particles within 500 pc, where approximately 75 percent of the tracers
reside at the start of the simulations. As Fig.~\ref{fig:M<500pc}
shows, this decline occurs on the relaxation time scale $t_{\rm rh}$,
and may be explained by the energy exchange between massive particles
within the half-mass radius.  The inward flow of heat leads to an
expansion of the inner halo and a decrease in the central density.
This is a manifestation of the inverted temperature profile of the NFW
model (see Fig.~\ref{fig:vdisp}), as compared with the more usually
considered Plummer model.  We note that such an inversion is not
unique to the NFW model, and other profiles exist which also display
such an effect \citep[e.g.][]{dehnen03}.  The analytic dispersion
shows that the massive particles at a radius of approximately $r_{\rm
  s}$ are initially hotter than those in both the inner and outer
regions.  Following this evolution towards an isothermal velocity
distribution, the subsequent increase in the number of massive
particles within the same region, due to the well understood negative
specific heat of self-gravitating systems, may then be described as a
slow evolution towards core collapse, similar to what is found in the
Plummer model ~\citep{Cohn}.  As was mentioned earlier, this evolution
was also noted in the simulations of~\cite{Hayashi2003}.

\subsection{Stellar Heating Mechanism}
\label{sec:equipartition}
\begin{figure}
  \resizebox{0.5\textwidth}{!}{
\includegraphics[]{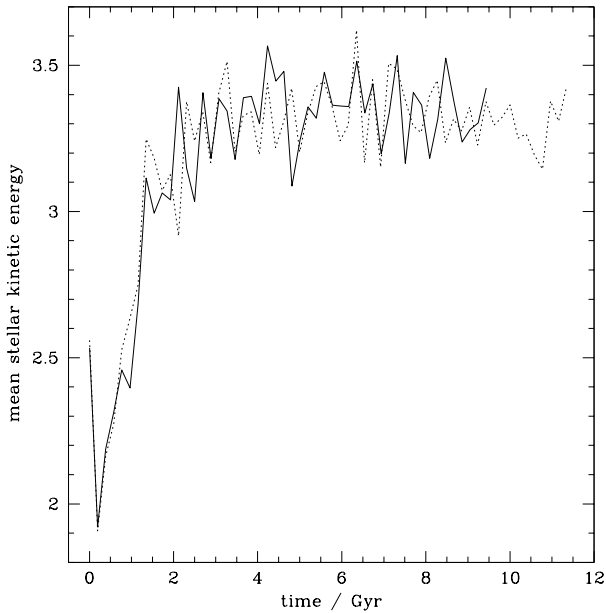}}
\caption{The mean tracer kinetic energy (in $10^{-10}$ $N$-body units) as a
function of time for models 4 (solid) and 6 (dotted).  Evolution
towards equipartition rapidly results in the tracers acquiring energy
from the black holes.\label{fig:energy}}
\end{figure}

The other main effect observed in our simulations is illustrated in
Fig.~\ref{fig:trcount}, namely the rapid decline in tracer numbers
within 500 pc.  As the figure shows, reducing the mass of the black
holes comprising the halo decreases the rate at which this heating
occurs, indicating that the mechanism for the increase in mean stellar
kinetic energy is equipartition with the massive particles.

If a system consists of multiple populations with different particle
masses, then encounters will tend to lead to the establishment of
equipartition of kinetic energies, with the more massive particles
losing energy to the less massive particles over time, assuming that
velocity was initially independent of mass.  As a consequence, the
more massive particles will gradually lose kinetic energy and move
towards the system centre, whereas the less massive particles will
move to larger orbits.  The time scale on which equipartition occurs is
related to the half-mass relaxation time.  In order to check that the
rate of heating of the tracers is consistent with kinetic energy
equipartition, we calculate the theoretical rate of energy transfer
between the two populations given their mean kinetic energies at a
specified time. This is given by~\citep[ eq. 2-60]{Spitzer}:

\begin{equation}
\label{eq:Spitzer}
  \frac{\ud\bar{E}_{t}}{\ud t} =
  2\left(\frac{6}{\pi}\right)^{1/2}\frac{m_t n_b
  \Gamma}{m_b}\frac{(\bar E_b - \bar E_t)}{(v_{t}^2 + v_{b}^2)^{3/2}}\, ,
\end{equation} 
where $m$ is the mass of an individual particle, $n$ is the average
number density of the black holes, $\bar E$ is the mean particle
kinetic energy given by

\begin{equation}
  \bar E_i = \frac{1}{2}m_{i} v_{i}^2\, , 
\end{equation}
and $\Gamma$ is defined by

\begin{equation}
   \Gamma \equiv 4\pi G^2 m_{b}^2 \,\mathrm{ln}\Lambda \,.
\end{equation} 
The subscripts $t$ and $b$ refer to tracer and black hole
respectively.  Since the halo is observed to evolve on the relaxation
time scale at the half-mass radius $r_{\rm h}$, we take $n$ to be the
mean number density within $r_{\rm h}$.  The value of $\Lambda$ in the
Coulomb logarithm is given by $\Lambda = 0.4N$ where $N$ is the total
number of particles of both populations within $r_{\rm h}$.

Equation~(\ref{eq:Spitzer}) can be recast in a more transparent
fashion by defining $\bar \varepsilon$, the energy per unit mass of
the relevant particle species, and replacing $n_b m_b$ by $\rho_b$. If
we consider the limit where $\bar E_b \gg \bar E_t$, then we find

\begin{equation}
\label{eq:Spitzer_mod}
  \frac{\ud\bar{\varepsilon}_{t}}{\ud t} =
  4\left(6\pi\right)^{1/2}\frac{v_b^2}{(v_b^2 + v_t^2)^{3/2}}\,G^2
  \rho_b \,m_b \,\mathrm{ln}\Lambda \,.
\end{equation} 
For a fixed mass density in the gravitationally dominant species, the
rate of approach to equipartition increases in proportion to the black
hole mass and decreases as the cube of the velocity of the tracer
particles.

The evolution of tracer energy is shown in Fig.~\ref{fig:energy} for
models 4 and 6, where the rate of energy change is observed to be
steady for approximately 2 Gyr.  For model 4, the difference in the
mean stellar kinetic energy in $N$-body units is approximately
$1.12\times 10^{-10}$ between 0.2 and 2 Gyr which corresponds to 10
$N$-body time units.  The rate of energy transfer according to
equation~(\ref{eq:Spitzer}) varies as a function of time, but is
$\sim1.5$$\times 10^{-11}$ at 0.2~Gyr and $\sim$$1.1\times 10^{-11}$
at 2 Gyr.  Based on these rates we expect to observe an energy
transfer in the range $(1.1\sim1.5)\times 10^{-10}$ units, which
agrees well with the energy difference between these two times.

\begin{figure}
  \resizebox{0.5\textwidth}{!}{
\includegraphics[]{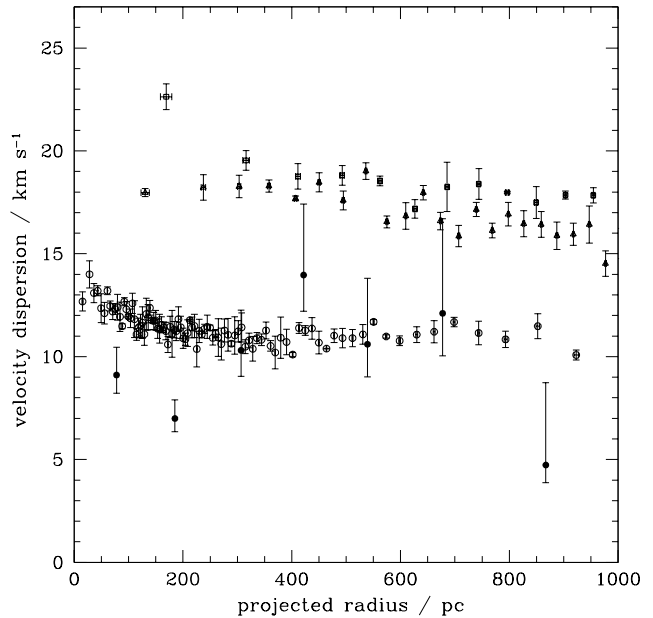}}
\caption{One dimensional (line-of-sight) velocity dispersions of model 4 at
  times $T = 0$, 10 and 20 Gyr (shown with open circles, open
  triangles and open squares respectively) overplotted with Draco
  velocity dispersion profile from 
~\citet{Mark2004} 
(filled circles,
  with $1\sigma$ errorbars).\label{fig:trvdisp-oplot}}
\end{figure}
\begin{figure}
  \resizebox{0.5\textwidth}{!}{
\includegraphics[]{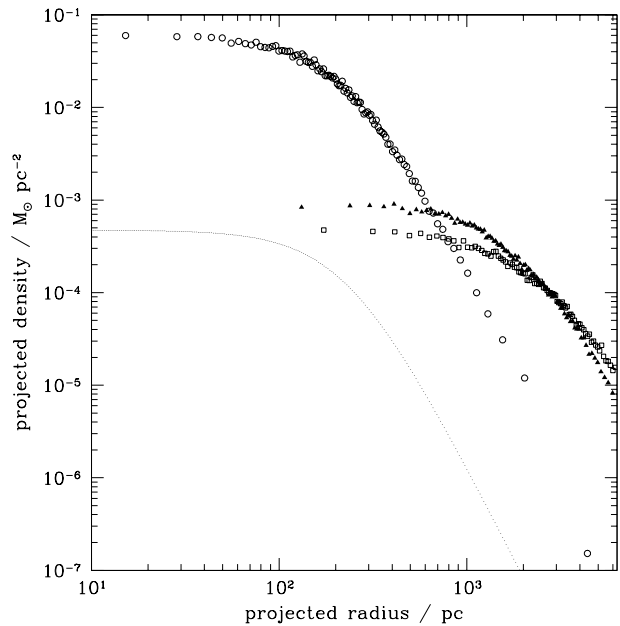}}
\caption{Overplot of the tracer density in projection for model 4 and the
Plummer profile for Draco.  Projected densities at times $T = 0$ (open
circles), 10 (filled triangles) and 20 (open squares) Gyr are shown.
The fit to the light distribution of Draco, scaled such that the
central density matches the simulated density at 20 Gyr, is given by
the dotted line.\label{fig:trdens-oplot}}
\end{figure}

Equation~(\ref{eq:Spitzer_mod}) shows that the rate at which the stars
gain energy from a population of black holes of mass $m_{\rm b}$ and mean
spatial density $\rho_{\rm b}$ is only weakly dependent on the details of
the mass distribution. In order to confirm this, we have performed a number of
simulations (not presented in Table~\ref{tab:sim}) in which the halo is
represented by a Plummer sphere whose mean density within 3 stellar core
radii matches that of the NFW profile in model~4.  These simulations
confirm that the stellar heating rates for both cored and cusped black
hole haloes are almost identical for haloes with similar mean density in
the region initially occupied by the stellar distribution.

The formulae given above are based on the assumption that the change
in energy of a test particle is due to the cumulative effect of many
weak encounters rather than a single close encounter.  It is therefore
necessary to be aware of the frequency with which tracers experience
close encounters with the black holes that may render the use of
equation~(\ref{eq:Spitzer}) invalid.  For models 4 and 6, we find that
such encounters are relatively rare, and the results of our energy
calculations above confirm that the gradual transfer of energy to the
tracers by the black holes initially within $r_{\rm h}$ is the main
mechanism by which energy is transferred between the two
populations. This may be partly due to the value of the softening
parameter used in those models which tends to suppress large-angle
scattering events.

\subsection{The Effect of Force Softening}
\label{subsec:grape}

Hard black hole-black hole binaries may have an important impact on
the dynamics of the entire halo.  Softening in turn is a key factor in
binary dynamics, from the rate of formation to their rate of
destruction.  We have re-simulated models 4 and 11 using the NBODY4
code on a GRAPE-6, with the aim of quantifying the effects of a
smoothed force-law.  Each of the five re-simulations were performed
using a different random seed to compute the initial conditions; apart
from the exclusion of tracers, all other parameters were kept the same
as the original models.  Hard binaries (i.e. those with $|E_{\rm bin}|
> m \sigma^2$ where $m$ is the mean mass and $\sigma$ is the velocity
dispersion) are identified in these simulations and treated using
regularisation, to enable an accurate and efficient treatment of their
internal motion \citep{Aarseth:99,ma93,ma98}.  The results confirmed
that the presence of binaries is a very important factor in
determining the dynamics of the remainder of the system, with a single
binary in the central region being sufficient to cause significant
fluctuations in the central density.

In the simulations with $10^6\Msun$ black holes, we observe a
relatively steady rate of binary formation and destruction.  These
binaries tend to form within the inner $1\kpc$ region and often
undergo several companion swaps before being destroyed or expelled.
Although this has interesting consequences for the evolution of the
inner region of the halo in particular, the fact that we do not
observe any binaries in simulations with $10^5\Msun$ black holes,
coupled with the discussion below that black hole masses of greater
than $10^5\Msun$ are inconsistent with the preservation of the stellar
population in Draco, implies that our overall conclusions based on the
simulations with softening are unaffected by the inclusion of
softening.

\subsection{Comparison with the Draco dSph}
\label{subsec:draco}

\begin{figure}
  \resizebox{0.5\textwidth}{!}{
\includegraphics[]{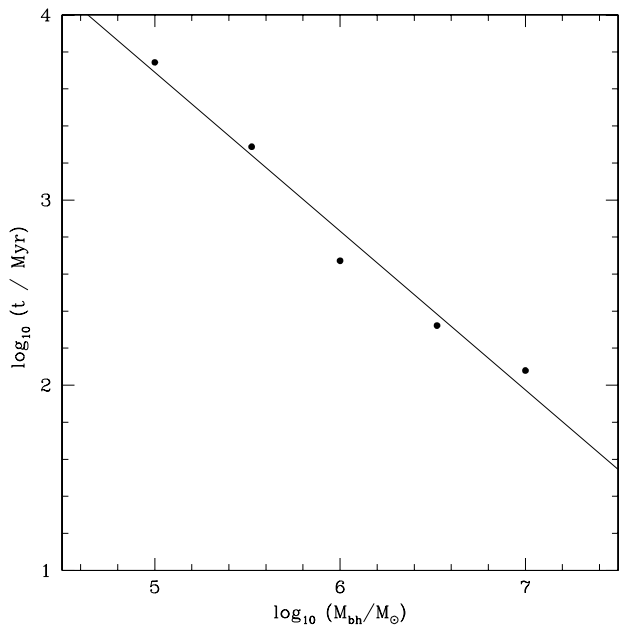}}
\caption{Dependence on the individual black hole mass of the time taken for the
initial tracer count within $500\pc$ to drop by a factor of 2.  The
points show this time for models 2--4, 8 and 11; the curve shows the fit
described by equation~(\ref{eq:fit}).\label{fig:ndrop}}
\end{figure}

A comparison of the observed and simulated data can be made by means
of the velocity dispersion and projected density shown in
Figs.~\ref{fig:trvdisp-oplot} and \ref{fig:trdens-oplot} respectively.
The former shows the velocity dispersion profile of Draco overplotted
on the dispersions for model 4 at three sampling times, while the latter
gives the light distribution at the same times. Also plotted in
Fig.~\ref{fig:trdens-oplot} is a Plummer profile, which is known to
provide a good fit to the light distribution in the inner region of
Draco. The profile has been scaled such that the central density
matches that of the core density in the simulation at 20 Gyr.  The
final model density profile in projection is clearly too extended to
be compatible with Draco's observed stellar distribution.

If the halo of Draco is indeed composed of massive black holes, the
current extent of the stellar distribution can be used to place an
upper limit on the mass of individual black holes, if we assume that
there is no spread in the mass of black holes in a halo. In
Fig.~\ref{fig:ndrop} we plot the time taken for the initial tracer
count within $500\pc$ to drop by a factor of 2 for models 2--4, 8 and
11. We would expect that this time $t$ would be related to $t_{\rm
rh}$ for the models and should therefore scale with mass approximately
as $M^{-0.8}$, in the mass range we are investigating. In fact, the
relation
\begin{equation}
\label{eq:fit}
  t = \frac{9.549\times 10^7}{M^{0.858}},
\end{equation}
with $t$ in Myr and $M$ in $\Msun$, provides a good fit to the results
from the simulations as shown in Fig.~\ref{fig:ndrop}.  If we require
that the process described above takes at least 5 Gyr, we are able to
set an upper limit on $M$ of $9.7\times 10^4\Msun$; for 10 Gyr, an
upper limit of $4.4\times 10^4\Msun$ would be implied. It should be
noted that a drop in the central tracer count by a factor of 2 often
corresponds to an increase in the scale radius of the tracer
population by a factor of a few and therefore these upper limits are
probably overestimates. Thus, black hole masses of less than
$10^5\Msun$ are required to prevent significant evolution of the
tracer density in a Hubble time; if Draco's initial stellar
distribution was similar to that observed today then black hole masses
in excess of $10^5\Msun$ can immediately be ruled out.

It is worth noting that black hole masses below $10^5\Msun$ are a
priori less interesting as dark matter candidates for a number of
reasons. First, as LO85 have shown, black hole masses of approximately
$10^6\Msun$ are required to explain the heating of the Milky Way
disk. Second, haloes comprising lower mass black holes evolve
significantly more slowly than haloes of more massive holes and a halo
which formed with a dark matter cusp would, therefore, not develop a
core in less than a Hubble time. We therefore conclude that if it
could be demonstrated that Draco's stellar distribution has not
changed significantly over a Hubble time, then the two primary
motivations for adopting black holes as dark matter candidates would
disappear.

Another consideration is whether our restricted choice of initial
conditions would affect our black hole mass limits.  We have made a
number of simplifying assumptions in our choice of initial conditions.
First we assume spherical symmetry for our halo distributions.  This
is a reasonable starting point since dSphs such as Draco are observed
to have an ellipticity of the stellar distribution of only
$\epsilon_\rho \simeq 0.3$ \citep{Mateo1998}.  The ellipticity of the
underlying potential is typically about 1/3 that of the density
\citep[e.g.][]{BT87} and the spherical symmetry is probably a good
approximation to Draco.  Second, we assume that the dSph evolves in
isolation -- as discussed earlier, observations of Draco suggest that
it has been only weakly affected by the tidal field of the Milky Way.
Third, we have assumed that all our black holes have the same mass --
this effectively means that we are assuming a strongly peaked black
hole mass function.  The key point with regard to our models is that
including any of the above effects e.g. a spectrum of black hole
masses or an external tidal field would serve to speed up the
evolution of the black hole halo \citep{giersz97}.  Our simulations
therefore represent a lower limit to the expected rate of halo
evolution.  The inclusion of a spectrum of black hole masses could
reduce the time to core collapse to $\sim 1 t_{\rm r}$
~\citep{giersz96}. From Table~1, allowing for the initial evolution of
the NFW profiles towards a core, models in which the peak of the black
hole mass function is less than about $10^{5.5}$M$_\odot$ are still
unlikely to reach core collapse within a Hubble time. On the other
hand, models with a mass peak around $10^6$M$_\odot$ or above will
almost certainly achieve core collapse. This adds further weight to
our conclusion that black holes of $10^6$M$_\odot$ and above are
unlikely dark matter candidates.

\section{Compact Initial Conditions}
\label{sec:init_cond}
\begin{figure}
  \resizebox{0.5\textwidth}{!}{
\includegraphics[]{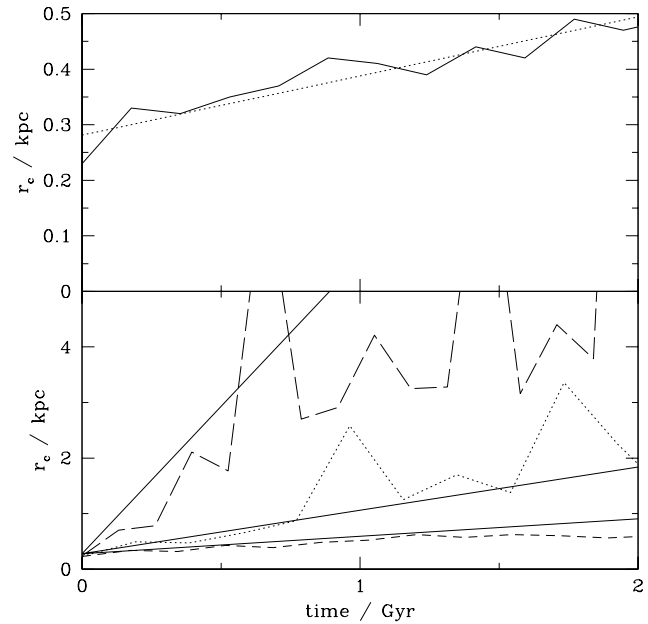}}
\caption{Top panel: growth of the scale radius of the tracer distribution
for model 11 as a function of time (solid line). The dashed line shows
the straight line fit used to describe this growth (see text for
details). Bottom panel: tracer scale radius evolution for models 2
(long-dashed curve), 4 (dotted curve) and 8 (short dashed) together
with predicted growth rates (solid lines) based on a rescaling of the
fitted curve for model 11.}
\label{fig:growth_fig}
\end{figure}

The results of the previous section demonstrate that the present
stellar distribution of Draco can be preserved for a Hubble time only
for black hole masses less than $10^5\Msun$. In this section we
consider an alternative possibility, namely that the stellar
distribution was initially significantly more compact than it is at
present and was subsequently heated to its current condition by the
processes described above. We now present a simple scaling argument
which demonstrates that the initial conditions required are not
unreasonable for black hole masses of about $10^5\Msun$.

In order to quantify the evolution of the tracer population, we fit
three dimensional density profiles of the form~\citep{Zhao1996}
\begin{equation}
\rho(r) = \frac{\rho_0 r_{\rm c}^\gamma}{r^\alpha\left(r^\beta 
+ r_{\rm c}^\beta\right)^{(\gamma-\alpha)/\beta}}.
\end{equation}
These profiles incorporate the Plummer profile ($\alpha=0$, $\beta=2$,
$\gamma = 5$) but they also include outer profiles which fall off
either faster or slower than $r^{-5}$. The curves in
Fig.~\ref{fig:growth_fig} show the evolution of the tracer scale radii
$r_{\rm c}$ obtained from fits of this form to four of our models.

The top panel of Fig.~\ref{fig:growth_fig} shows the evolution of the
scale radius of the tracer distribution in model 11 during the first 2
Gyr. Until this time, more than $70$ percent of the tracers lie within
one scale radius $r_{\rm s}$ of the halo. This is important, as we
expect the simple scaling arguments given below to hold only during
the time when the tracers are confined to the central regions of the
halo, where the density profile is a single power-law. The solid line
in Fig.~\ref{fig:growth_fig} shows the straight line fit to the
temporal increase of $r_{\rm c}$
\begin{equation}
\label{eqn:heating_rate}
r_{\rm c} (t) = 0.28 \left( 1 + 9.467 \left( \frac{t}{t_{\rm
rh}}\right) \right),
\end{equation}
where $r_{\rm c}$ is in kpc. In the regime where the majority of
tracers lie in the volume where the halo potential goes as $1/r$, the
parameters of the fit should be independent of the mass of the black
holes and the evolution of the scale radius can therefore be estimated
from the above equation using the appropriate value for $t_{\rm
rh}$. For the other models, this approximation is valid for less than
$2\Gyr$ as the tracer distribution rapidly expands into the outer
regions of the halo. However, as the bottom panel of
Fig.~\ref{fig:growth_fig} demonstrates, this simple scaling argument
yields a reasonable estimate of the initial growth rate of the tracer
scale radius. At later times, the heating rate falls below the
estimated value as the tracers no longer lie solely in the single
power-law region of the halo.

We can extrapolate the scale radius evolution estimated from
equation~(\ref{eqn:heating_rate}) to consider the evolution of tracer
populations whose initial distribution is more compact than those
considered so far. This extrapolation backwards in time is valid
unless the initial $r_{\rm c}$ is so small that the stellar population
would become self-gravitating. For example, for model 2, if we require
that after $10\Gyr$ we have $r_{\rm c} = 232\pc$, then this implies an
initial radius of $1\pc$ for the stellar population, which corresponds
to a dense and self-gravitating stellar cluster. However, for model 11
the required initial radius is $48\pc$. If this population resided in
the centre of the halo, then the halo mass within its volume would be
comparable to the stellar mass which would not, therefore, be
self-gravitating. This suggests that the heating of the stellar
population by a halo of $10^5\Msun$ black holes need not be
catastrophic for a dSph such as Draco, provided that its initial
stellar population is sufficiently compact. It is less clear whether
the same holds for black hole masses of $10^6\Msun$. In this case, the
initial cluster has scale radius $8\pc$ and since it would therefore be
almost self-gravitating, the simple scaling argument presented here
breaks down.

It is important to emphasise at this point that the heating rate
depends on the relative velocities of the black holes and tracers. A
more compact, but not self-gravitating, initial tracer density
distribution has a lower velocity dispersion and hence the rate at
which it is heated is initially greater than we have estimated
here. However, this increases the initial heating rate by at most a
factor of two for the range of initial conditions we consider and so
will not substantially affect our conclusions. There are additional
heating effects to be considered for an initially small system, such
as relaxation due to $\sqrt{N}$ fluctuations in the number of black
holes within the stellar distribution. These would make early
expansion proceed more rapidly than otherwise. While it might be
possible to balance this effect to some extent by the stabilising
effects of self-gravity, we still conclude that constructing a
dynamical history for Draco which results in a present-day stellar
distribution with a half-light radius of $232\pc$ may be problematic
for the scenario considered here. Further work is needed to establish
the exact degree of fine tuning of the initial conditions which would
be required.

If the stellar population of Draco began in a compact state and its
current extent is the result of heating by black holes, it is natural
to ask whether this process would lead to any clear observational
signatures in either the stellar luminosity or velocity distributions.
\cite{Spitzer} has shown that weak encounters in $N$-body systems
produce a population of weakly bound objects with a power-law density
distribution. The logarithmic slope of this density profile depends on
the gravitational potential, which is assumed also to be a single
power-law. LO85 showed that strong encounters generate a profile with
a different logarithmic slope to that generated by weak encounters. In
our simulations, however, both weak and strong encounters with the
black holes contribute to the heating of the stellar distribution. As
a result, there is no simple expression for the outer slope of the
resultant light distribution. If the tail generated by close
encounters were sufficiently populated by stars, and could therefore
be identified in the light distribution of Draco, it would argue
strongly that the stellar population had been heated through close
encounters. On the other hand, our results demonstrate that the
absence of such a tail does not preclude the black hole model. For
example, in model 5 the outer profile of the stellar distribution
after 10 Gyr goes approximately as $r^{-4.6}$ which would be
indistinguishable from a Plummer law, given the magnitude of the error
bars on the observed luminosity at such
radii~\citep[e.g.][]{Odenkirchen}.  We note that the change in the
slope of the light distribution in the outer parts of Draco recently
identified from deep imaging \citep{Mark2004} might be indicative of
such a power law halo.

The velocity distribution which arises in the black hole halo scenario
is also difficult to identify unambiguously from observations.  One
would expect to observe a significant population of stars on radial
orbits in the outer regions of a heated model as a result of the
heating process. Indeed, after 10 Gyr, the anisotropy of the velocity
distribution in model 5 is strongly radial in the outer
parts. However, the corresponding projected velocity distribution in
that model does not display any features which would unambiguously
point to the strong radial bias in the velocity distribution. We
conclude that the simulations presented here do not predict a unique
observable signature of heating by black holes. In order to address
this rather unsatisfactory situation, we will investigate the
evolution of initially compact dSphs in a future paper.

In our simulations we have shown that a halo which is composed of
massive black holes and which initially has an NFW density profile can
be transformed into a cored halo through dynamical evolution. A number
of authors have considered the role of small numbers of black holes in
transforming the density profiles of cusped stellar systems into
cores~\citep[e.g.][]{hemsendorf03} or cusped haloes into cored
haloes~\citep[e.g.][]{mm02}. In these papers, the black holes were a
minor constituent of the mass distribution, contrary to the case we
consider in which the black holes are the dark matter and constitute
the bulk of the gravitating mass. We are not aware of any simulation
of the evolution of a compact stellar system in the presence of an
extended halo of black holes. Of more direct relevance to our results,
we note that ~\cite{hg98} concluded that black holes with masses above
$10^5\Msun$ could not constitute more than $1/8$ of the total mass
of dwarf disk galaxy haloes, as in this case dynamical friction would
lead to a concentration of mass at the centre of the dwarf and a
consequent observable signature in the rotation curve of the
dwarf. Our results cannot be compared directly with those of
~\citeauthor{hg98} as their calculation assumes a smooth background
halo distribution with a uniform density core, while in our models no
such smooth background exists. However, it is interesting that
~\citeauthor{hg98} came to similar conclusions about the
upper limit to the mass of any putative black hole population.

We conclude this section with a discussion of potential criticisms of
our black hole halo model. First, one might argue that the probability
that we would observe a significant number of dSphs at a similar phase
of their evolution should be very small, which would render our
compact initial conditions scenario implausible. This objection can be
countered if we assume that dwarf elliptical galaxies and dSphs were
originally members of the same class. Dwarf ellipticals have high
stellar densities compared with the low densities of the dSphs, and
observations have produced evidence for the presence of black holes in
these galaxies with masses of a few million solar masses in a number
of these systems~\citep[e.g. M32:][]{Verolme02}. As we have seen,
heating of a compact stellar system by a black hole halo could produce
a low surface brightness object similar to a dSph. On the other hand,
in some cases a single black hole might become embedded in the compact
cluster protecting it from further shredding and causing the stellar
density to remain high, as in M32.  If there is a characteristic black
hole mass, then the evolutionary time scales of all the dSphs would be
similar and hence it is not surprising that we observe them today in
similar evolutionary states.

A second objection to the hypothesis that black holes constitute the
dark matter in dSph galaxy haloes arises from observations of the
Fornax dSph. Fornax, the most luminous Local Group dSph, is surrounded
by a population of five old, globular clusters. The presence of these
clusters in an extended distribution about the galaxy is difficult to
reconcile with the presence of a large number of black holes whose
masses are comparable to those of the clusters, because close
encounters between the clusters and the black holes would be expected
to lead to the rapid disruption of the clusters. While this is indeed
perplexing, we note that if the dark matter is composed of much lower
mass particles (e.g. subatomic particles), then the extended radial
distribution of the Fornax clusters is equally surprising, as
dynamical friction should draw the clusters into the centre of the
galaxy in a fraction of the Hubble
time~\citep[e.g.][]{tos75,olr00}. Thus the origin of the Fornax
globular clusters remains an open question and, in the absence of
further data, does not provide a means of distinguishing between the
conventional model for the dark matter and the proposal considered
here.

There are also two sets of objects in the Milky Way halo, observations
of which appear to argue against black holes as major constituents of
the dark matter. If black holes were ruled out for the Milky Way halo,
it would make it less likely that the dark matter in dSph galaxies
could be composed of black holes, since disruption of dSphs over the
lifetime of the Milky Way would pollute its halo with black holes.
First, deep imaging of the Milky Way globular cluster Pal~5 has
revealed the presence of long tidal tails~\citep{odenk01,odenk03}
extending to distances of about $2$ kpc from the cluster. Estimates of
the drift rate of stars along these tails suggest that the time
required for the tails to reach their observed extent is approximately
$2$ Gyr~\citep{odenk03}. It has been pointed out that the spatial
integrity of the streams, which implies that they are kinematically
cold, may place constraints on the properties of any massive
substructures present in the halo of the Milky
Way~\citep[e.g.][]{dehnen04}. In particular, simulations by Moore
(priv. comm.) suggest that the tails of Pal~5 rule out a significant
population of black holes with masses of about $10^6\Msun$ in the halo
of the Milky Way. However, we note that similar arguments apply to the
substructure which is ubiquitous in standard cold dark matter
simulations and thus the tails of Pal~5 do not constitute a clear
observational test of the black hole model. We also note
that~\cite{dehnen04} have shown that $N$-body simulations of the tidal
disruption of Pal~5 in a smooth Milky Way halo are unable to reproduce
the observed internal structure in the stellar density distribution
along the tails. Those authors suggest that the presence of an amount
of halo substructure may be required to account for the non-uniform
distribution of stars along the tails of Pal~5.

Recent work on the distribution of separations of wide stellar
binaries in the halo of the Milky Way may constitute the strongest
observational argument against the black hole halo model. In a
comprehensive study of this issue,~\cite{ycg04} show that encounters
between wide binaries and massive compact objects would result in the
power-law tail of binary separations becoming steeper with time. They
argue that the absence of a cut-off in the distribution of wide binary
separations in a large sample of nearby halo binaries rules out the
presence of compact objects with masses above $43\Msun$ at the
standard local halo density. This analysis assumes that all the
binaries in their sample have been part of the Galactic halo for at
least 10 Gyr and during that time have experienced a density of dark
matter comparable to the local halo density. Given that most of the
halo binaries were probably formed in star clusters, it is not clear
how long they have been exposed directly to potentially disrupting
encounters with halo substructure. Further, if the binaries originated
in halo star clusters on highly elongated orbits, they may have spent
much of their lives in regions where the dark matter density is
significantly lower than in the solar neighbourhood. We conclude that
without information about the actual distribution of ages for the halo
binaries (i.e. how long they have been part of the halo) and knowledge
of the orbital distribution of the binaries in the Galactic halo, the
population of wide binaries is not yet sufficient to rule out
definitively massive objects as a significant component of the dark
matter. However, these data have the potential to rule out black hole
dark matter and therefore merit further analysis to determine their
orbital properties.

\section{Summary and Conclusions}
\label{sec:con}
The high mass-to-light ratios of certain Local Group dwarf galaxies
suggest that they are among the most dark matter-dominated stellar
systems known in the Universe. This makes them ideal laboratories in
which to investigate the nature and properties of dark matter.  In
addition, their relative proximity has allowed the accumulation of a
wealth of high quality stellar kinematic data in recent years.  We
present here the results of simulations where the modelled system is
assumed to consist of a dwarf galaxy-sized stellar population, which
initially resides in the centre of an extended, massive halo with an
NFW profile.  The halo particles are assigned masses of between
$10^{5}\Msun$ and $10^7\Msun$ and are modelled as massive black holes.
The mass of the system is chosen to match that of the Draco dSph, a
Galactic satellite for which evidence has recently been presented of
the existence of a massive dark matter halo ~\citep{Jan2001}.  The
stellar distribution appears to have suffered only minor tidal
distortion ~\citep{Klessen2003,Mark2004}. This, together with the
availability of stellar velocity measurements up to four Plummer radii
\citep{Jan2002,Mark2004}, makes Draco a useful test candidate for
comparison with simulations.

The first question which our simulations address is whether the black
hole population evolves too quickly to be consistent with the
potential observed for Draco.  Perusal of the very short central
relaxation times (Table~\ref{tab:sim}) indicates that we might have
expected the systems to undergo core collapse and re-expansion within
a Hubble time, with subsequent dissolution in a Galactic tidal field,
had the initial conditions been closer to a Plummer model
~\citep{Lee1987}.  However, due to the inverted initial temperature
profile of the NFW model, the evolution in our simulations is slower
and initially in the opposite direction, towards a lower central
density with a cored rather than a cusped density profile.  Only
systems with black hole mass greater than $10^{6.5} \Msun$ appear to
evolve too rapidly towards core collapse.  For the lower end of the
mass range investigated, evolution leads to a profile with a density
profile that is more consistent with observations of dSphs
~\citep{Jan2003,Magorrian2003} than the initial NFW model.

A more serious set of issues is raised by the evolution of the tracer
populations.  We find that in all of our dSph-sized models, the
central tracer density shows a very rapid decline in the first Gyr,
corresponding to over-efficient heating by the black holes.  This
effect can also be observed via the increase in the tracer velocity
dispersion, which at 10 Gyr cannot be reconciled with the observed
Draco dispersion. If the initial stellar distribution is chosen to
match that of Draco, then only for black holes of mass below about
$10^5\Msun$ is the heating rate reduced to a level which could perhaps be
compatible with the observations.  We conclude that the haloes of dSph
galaxies cannot be composed of black holes more massive than
$10^5\Msun$ if their initial stellar distributions were similar to
those observed at the present time.

We have also performed simulations in which the scale length was set
to twice that used to model the halo of Draco.  The heating of the
stellar population in these cases was somewhat reduced, and more than
$10$ percent of the stellar population which initially occupied the
central $500\pc$ of the halo remained in this region after 10 Gyr for
black hole masses of $10^6 \,\textrm{M}_{\odot}$, as opposed to a loss
of greater than 95 percent for the models with smaller $r_{\rm s}$.
However, the velocity dispersion of the evolved tracers within the
same region is still uncomfortably large. Notwithstanding the question
of how to suppress the heating of the stellar population, the
evolution of the haloes of these models is potentially very
interesting. As Fig.~\ref{fig:vcirc} shows, the inner regions of the
haloes evolve towards a more cored profile. Thus, these models may be
able to explain the observations of LSB galaxies which appear to have
large dark matter cores~\citep{Blok2002}.

By constraining the amount of evolution allowed to take place for the
stellar population within a given time scale, we are able to place a
new upper limit on the mass of individual black holes which constitute
the dark matter halo of Draco. If we require that the stellar count
within $500\pc$ of the halo origin decreases by no more than a factor
of 2 within 5 Gyr, we obtain an upper mass constraint of approximately
$10^{5}\Msun$; the equivalent condition required over a time scale of
10 Gyr sets a limit of $10^{4.6}\Msun$. These numbers call into
question the hypothesis that massive black holes could be a
significant component of the dark matter: black holes with masses
below $10^5\Msun$ are less interesting as dark matter candidates
because they cannot explain either the heating of the Milky Way disk
(LO85) or the origin of cores in dark haloes
(Section~\ref{subsec:draco}). Therefore, we have also investigated the
possibility that a dSph whose initial stellar distribution was more
compact than those of the present-day dSphs could evolve into an
object resembling Draco as a result of heating. We have presented a
simple scaling argument which implies that if the black holes have
masses of $10^5\Msun$ then an initial stellar distribution with scale
radius $48\pc$ would evolve into an object of similar extent to that
of Draco in $10\Gyr$; such a scenario may permit models with black
hole masses somewhat larger than $10^5\Msun$ and correspondingly more
compact initial conditions. However, we note that this scenario may
require very finely tuned initial conditions because $\sqrt{N}$ noise
in the black hole distribution may inflate the heating rate at early
times.

In summary, we find that scenarios in which the dark matter is in the
form of massive black holes, with masses between $10^5\Msun$ and a few
$\times 10^5\Msun$, may provide a viable channel for the production of
cored, rather than cusped, haloes but only if the initial stellar
distributions of the dSphs were considerably more compact than the
currently observed ones. Further simulations will be required to
determine if this picture is, in all details, consistent with the
observed data on the Local Group dSphs.

\subsection*{Acknowledgements}
We thank Scott Tremaine for valuable comments on an earlier draft of
this paper and Sverre Aarseth for many useful discussions and for the
use of his NBODY codes. We thank Andy Gould, Bohdan Paczynski, Piotr
Popowski, Paolo Salucci and F. Javier Sanchez for their critical
comments on the feasibility of the models, and the anonymous referee
for useful comments.  MIW acknowledges financial support from PPARC.


\begin{thebibliography}{}

\bibitem[\protect\citeauthoryear{{Aarseth}}{{Aarseth}}{1999}]{Aarseth:99}
{Aarseth} S.~J.,  1999, Celestial Mechanics and Dynamical Astronomy, 73, 127

\bibitem[\protect\citeauthoryear{{Aarseth}}{{Aarseth}}{2001}]{Sverre}
{Aarseth} S.~J.,  2001, New Astronomy, 6, 277

\bibitem[\protect\citeauthoryear{{Binney} \& {Tremaine}}{{Binney} \&
  {Tremaine}}{1987}]{BT87}
{Binney} J.,  {Tremaine} S.,  1987, {Galactic dynamics}.
Princeton, NJ, Princeton University Press

\bibitem[\protect\citeauthoryear{{Binney} \& {Evans}}{{Binney} \&
  {Evans}}{2001}]{binev01}
{Binney} J.~J.,  {Evans} N.~W.,  2001, \mnras, 327, L27

\bibitem[\protect\citeauthoryear{{Borriello}, {Salucci} \&
  {Danese}}{{Borriello} et~al.}{2003}]{bsd03}
{Borriello} A.,  {Salucci} P.,    {Danese} L.,  2003, \mnras, 341, 1109

\bibitem[\protect\citeauthoryear{{Bullock}, {Kolatt}, {Sigad}, {Somerville},
  {Kravtsov}, {Klypin}, {Primack} \& {Dekel}}{{Bullock}
  et~al.}{2001}]{Bullock2001}
{Bullock} J.~S.,  {Kolatt} T.~S.,  {Sigad} Y.,  {Somerville} R.~S.,  {Kravtsov}
  A.~V.,  {Klypin} A.~A.,  {Primack} J.~R.,    {Dekel} A.,  2001, \mnras, 321,
  559

\bibitem[\protect\citeauthoryear{{Cohn}}{{Cohn}}{1980}]{Cohn}
{Cohn} H.,  1980, \apj, 242, 765

\bibitem[\protect\citeauthoryear{{Colin}, {Klypin}, Valenzuela \&
  Gottlober}{{Colin} et~al.}{2003}]{Colin2003}
{Colin} P.,  {Klypin} A.,  Valenzuela O.,    Gottlober S.,  2003,
  astro-ph/0308348

\bibitem[\protect\citeauthoryear{{de Blok} \& {Bosma}}{{de Blok} \&
  {Bosma}}{2002}]{Blok2002}
{de Blok} W.~J.~G.,  {Bosma} A.,  2002, \aap, 385, 816

\bibitem[\protect\citeauthoryear{{De Lucia}, {Kauffmann}, {Springel}, {White},
  {Lanzoni}, {Stoehr}, {Tormen} \& {Yoshida}}{{De Lucia}
  et~al.}{2004}]{DeLucia03}
{De Lucia} G.,  {Kauffmann} G.,  {Springel} V.,  {White} S.~D.~M.,  {Lanzoni}
  B.,  {Stoehr} F.,  {Tormen} G.,    {Yoshida} N.,  2004, \mnras, 348, 333

\bibitem[\protect\citeauthoryear{{Dehnen}}{{Dehnen}}{2003}]{dehnen03}
{Dehnen} W., 2003, \mnras, 265, 250

\bibitem[\protect\citeauthoryear{{Dehnen}, {Odenkirchen}, {Grebel} \&
  {Rix}}{{Dehnen} et~al.}{2004}]{dehnen04}
{Dehnen} W.,  {Odenkirchen} M.,  {Grebel} E.~K.,    {Rix} H.,  2004, \aj, 127,
  2753

\bibitem[\protect\citeauthoryear{{Fukushige} \& {Makino}}{{Fukushige} \&
  {Makino}}{2001}]{Fukushige2001}
{Fukushige} T.,  {Makino} J.,  2001, \apj, 557, 533

\bibitem[\protect\citeauthoryear{{Ghigna}, {Moore}, {Governato}, {Lake},
  {Quinn} \& {Stadel}}{{Ghigna} et~al.}{2000}]{Ghigna2000}
{Ghigna} S.,  {Moore} B.,  {Governato} F.,  {Lake} G.,  {Quinn} T.,    {Stadel}
  J.,  2000, \apj, 544, 616

\bibitem[\protect\citeauthoryear{{Giersz} \& {Heggie}}{{Giersz} \&
  {Heggie}}{1996}]{giersz96}
{Giersz} M.,  {Heggie} D.~C.,  1996, \mnras, 279, 1037

\bibitem[\protect\citeauthoryear{{Giersz} \& {Heggie}}{{Giersz} \&
  {Heggie}}{1997}]{giersz97}
{Giersz} M.,  {Heggie} D.~C.,  1997, \mnras, 286, 709

\bibitem[\protect\citeauthoryear{{Gnedin} \& {Ostriker}}{{Gnedin} \&
  {Ostriker}}{2001}]{Gnedin_Ostriker}
{Gnedin} O.~Y.,  {Ostriker} J.~P.,  2001, \apj, 561, 61

\bibitem[\protect\citeauthoryear{{Hayashi}, {Navarro}, {Taylor}, {Stadel} \&
  {Quinn}}{{Hayashi} et~al.}{2003}]{Hayashi2003}
{Hayashi} E.,  {Navarro} J.~F.,  {Taylor} J.~E.,  {Stadel} J.,    {Quinn} T.,
  2003, \apj, 584, 541

\bibitem[\protect\citeauthoryear{{Heggie} \& {Mathieu}}{{Heggie} \&
  {Mathieu}}{1986}]{Heggie}
{Heggie} D.~C.,  {Mathieu} R.~D.,  1986, in LNP Vol. 267: The Use of
  Supercomputers in Stellar Dynamics p.~233

\bibitem[\protect\citeauthoryear{{Helmi} \& {White}}{{Helmi} \&
  {White}}{2001}]{Helmi2001}
{Helmi} A.,  {White} S.~D.~M.,  2001, \mnras, 323, 529

\bibitem[\protect\citeauthoryear{{Hemsendorf}}{{Hemsendorf}}{2003}]{hemsendorf%
03}
{Hemsendorf} M.,  2003, \apss, 284, 561

\bibitem[\protect\citeauthoryear{{Hernandez} \& {Gilmore}}{{Hernandez} \&
  {Gilmore}}{1998}]{hg98}
{Hernandez} X.,  {Gilmore} G.,  1998, \mnras, 297, 517

\bibitem[\protect\citeauthoryear{{Irwin} \& {Hatzidimitriou}}{{Irwin} \&
  {Hatzidimitriou}}{1995}]{IH95}
{Irwin} M.,  {Hatzidimitriou} D.,  1995, \mnras, 277, 1354

\bibitem[\protect\citeauthoryear{{Kazantzidis}, {Magorrian} \&
  {Moore}}{{Kazantzidis} et~al.}{2004}]{kmm04}
{Kazantzidis} S.,  {Magorrian} J.,    {Moore} B.,  2004, \apj, 601, 37

\bibitem[\protect\citeauthoryear{{Keeton}}{{Keeton}}{2003}]{Keeton2003}
{Keeton} C.~R.,  2003, \apj, 584, 664

\bibitem[\protect\citeauthoryear{{Klessen}, {Grebel} \& {Harbeck}}{{Klessen}
  et~al.}{2003}]{Klessen2003}
{Klessen} R.~S.,  {Grebel} E.~K.,    {Harbeck} D.,  2003, \apj, 589, 798

\bibitem[\protect\citeauthoryear{{Klessen} \& {Kroupa}}{{Klessen} \&
  {Kroupa}}{1998}]{Klessen_Kroupa}
{Klessen} R.~S.,  {Kroupa} P.,  1998, \apj, 498, 143

\bibitem[\protect\citeauthoryear{{Kleyna}, {Wilkinson}, {Evans}, {Gilmore} \&
  {Frayn}}{{Kleyna} et~al.}{2002}]{Jan2002}
{Kleyna} J.,  {Wilkinson} M.~I.,  {Evans} N.~W.,  {Gilmore} G.,    {Frayn} C.,
  2002, \mnras, 330, 792

\bibitem[\protect\citeauthoryear{{Kleyna}, {Wilkinson}, {Evans} \&
  {Gilmore}}{{Kleyna} et~al.}{2001}]{Jan2001}
{Kleyna} J.~T.,  {Wilkinson} M.~I.,  {Evans} N.~W.,    {Gilmore} G.,  2001,
  \apjl, 563, L115

\bibitem[\protect\citeauthoryear{{Kleyna}, {Wilkinson}, {Gilmore} \&
  {Evans}}{{Kleyna} et~al.}{2003}]{Jan2003}
{Kleyna} J.~T.,  {Wilkinson} M.~I.,  {Gilmore} G.,    {Evans} N.~W.,  2003,
  \apjl, 588, L21

\bibitem[\protect\citeauthoryear{{Lacey} \& {Ostriker}}{{Lacey} \&
  {Ostriker}}{1985}]{LO85}
{Lacey} C.~G.,  {Ostriker} J.~P.,  1985, \apj, 299, 633

\bibitem[\protect\citeauthoryear{{Lee} \& {Ostriker}}{{Lee} \&
  {Ostriker}}{1987}]{Lee1987}
{Lee} H.~M.,  {Ostriker} J.~P.,  1987, \apj, 322, 123

\bibitem[\protect\citeauthoryear{{Li} \& {Ostriker}}{{Li} \&
  {Ostriker}}{2002}]{Li2002}
{Li} L.,  {Ostriker} J.~P.,  2002, \apj, 566, 652

\bibitem[\protect\citeauthoryear{{Magorrian}}{{Magorrian}}{2003}]{Magorrian200%
3}
{Magorrian} J.,  2003, in The Mass of Galaxies at Low and High Redshift.
  Proceedings of the ESO Workshop held in Venice, Italy, 24-26 October 2001
  p.~18

\bibitem[\protect\citeauthoryear{{Makino}, {Taiji}, {Ebisuzaki} \&
  {Sugimoto}}{{Makino} et~al.}{1997}]{Makino:97}
{Makino} J.,  {Taiji} M.,  {Ebisuzaki} T.,    {Sugimoto} D.,  1997, \apj, 480,
  432

\bibitem[\protect\citeauthoryear{{Mao}, {Jing}, {Ostriker} \& {Weller}}{{Mao}
  et~al.}{2004}]{mao04}
{Mao} S.,  {Jing} Y.,  {Ostriker} J.~P.,    {Weller} J.,  2004, \apjl, 604, L5

\bibitem[\protect\citeauthoryear{{Mateo}}{{Mateo}}{1998}]{Mateo1998}
{Mateo} M.~L.,  1998, \araa, 36, 435

\bibitem[\protect\citeauthoryear{{Merritt} \& {Milosavljevi{\' c}}}{{Merritt}
  \& {Milosavljevi{\' c}}}{2002}]{mm02}
{Merritt} D.,  {Milosavljevi{\' c}} M.,  2002, in Dark matter in astro- and
  particle physics. Proceedings of the International Conference DARK 2002, H.
  V. Klapdor-Kleingrothaus, R. D. Viollier (eds.). Physics and astronomy online
  library. Berlin: Springer, 2002 p.~79

\bibitem[\protect\citeauthoryear{{Metcalf} \& {Madau}}{{Metcalf} \&
  {Madau}}{2001}]{Metcalf2001}
{Metcalf} R.~B.,  {Madau} P.,  2001, \apj, 563, 9

\bibitem[\protect\citeauthoryear{{Mikkola} \& {Aarseth}}{{Mikkola} \&
  {Aarseth}}{1993}]{ma93}
{Mikkola} S.,  {Aarseth} S.~J.,  1993, Celestial Mechanics and Dynamical
  Astronomy, 57, 439

\bibitem[\protect\citeauthoryear{{Mikkola} \& {Aarseth}}{{Mikkola} \&
  {Aarseth}}{1998}]{ma98}
{Mikkola} S.,  {Aarseth} S.~J.,  1998, New Astronomy, 3, 309

\bibitem[\protect\citeauthoryear{{Moore}, {Ghigna}, {Governato}, {Lake},
  {Quinn}, {Stadel} \& {Tozzi}}{{Moore} et~al.}{1999}]{Moore1999}
{Moore} B.,  {Ghigna} S.,  {Governato} F.,  {Lake} G.,  {Quinn} T.,  {Stadel}
  J.,    {Tozzi} P.,  1999, \apjl, 524, L19

\bibitem[\protect\citeauthoryear{{Navarro}, {Frenk} \& {White}}{{Navarro}
  et~al.}{1996}]{NFW96}
{Navarro} J.~F.,  {Frenk} C.~S.,    {White} S.~D.~M.,  1996, \apj, 462, 563

\bibitem[\protect\citeauthoryear{{Navarro}, {Frenk} \& {White}}{{Navarro}
  et~al.}{1997}]{NFW97}
{Navarro} J.~F.,  {Frenk} C.~S.,    {White} S.~D.~M.,  1997, \apj, 490, 493

\bibitem[\protect\citeauthoryear{{Odenkirchen}, {Grebel}, {Dehnen}, {Rix},
  {Yanny}, {Newberg}, {Rockosi}, {Mart{\'{\i}}nez-Delgado}, {Brinkmann} \&
  {Pier}}{{Odenkirchen} et~al.}{2003}]{odenk03}
{Odenkirchen} M.,  {Grebel} E.~K.,  {Dehnen} W.,  {Rix} H.,  {Yanny} B.,
  {Newberg} H.~J.,  {Rockosi} C.~M.,  {Mart{\'{\i}}nez-Delgado} D.,
  {Brinkmann} J.,    {Pier} J.~R.,  2003, \aj, 126, 2385

\bibitem[\protect\citeauthoryear{{Odenkirchen}, {Grebel}, {Harbeck}, {Dehnen},
  {Rix}, {Newberg}, {Yanny}, {Holtzman} \& {et al.}}{{Odenkirchen}
  et~al.}{2001a}]{Odenkirchen}
{Odenkirchen} M.,  {Grebel} E.~K.,  {Harbeck} D.,  {Dehnen} W.,  {Rix} H.,
  {Newberg} H.~J.,  {Yanny} B.,  {Holtzman} J.,    {et al.} 2001a, \aj

\bibitem[\protect\citeauthoryear{{Odenkirchen}, {Grebel}, {Rockosi}, {Dehnen},
  {Ibata}, {Rix}, {Stolte}, {Wolf} \& {et al.}}{{Odenkirchen}
  et~al.}{2001b}]{odenk01}
{Odenkirchen} M.,  {Grebel} E.~K.,  {Rockosi} C.~M.,  {Dehnen} W.,  {Ibata} R.,
   {Rix} H.,  {Stolte} A.,  {Wolf} C.,    {et al.} 2001b, \apjl

\bibitem[\protect\citeauthoryear{{Oh}, {Lin} \& {Richer}}{{Oh}
  et~al.}{2000}]{olr00}
{Oh} K.~S.,  {Lin} D.~N.~C.,    {Richer} H.~B.,  2000, \apj, 531, 727

\bibitem[\protect\citeauthoryear{{Ostriker} \& {Gnedin}}{{Ostriker} \&
  {Gnedin}}{1996}]{ostgne96}
{Ostriker} J.~P.,  {Gnedin} N.~Y.,  1996, \apjl, 472, L63

\bibitem[\protect\citeauthoryear{{Ostriker} \& {Steinhardt}}{{Ostriker} \&
  {Steinhardt}}{2003}]{Jerry2003}
{Ostriker} J.~P.,  {Steinhardt} P.,  2003, Science, 300, 1909

\bibitem[\protect\citeauthoryear{{Plummer}}{{Plummer}}{1911}]{Plummer1911}
{Plummer} H.~C.,  1911, \mnras, 71, 460

\bibitem[\protect\citeauthoryear{{Ricotti}}{{Ricotti}}{2003}]{Massimo2003}
{Ricotti} M.,  2003, \mnras, 344, 1237

\bibitem[\protect\citeauthoryear{{Salucci}}{{Salucci}}{2001}]{salucci01}
{Salucci} P.,  2001, \mnras, 320, L1

\bibitem[\protect\citeauthoryear{{Spergel} \& {Steinhardt}}{{Spergel} \&
  {Steinhardt}}{2000}]{SS2000}
{Spergel} D.~N.,  {Steinhardt} P.~J.,  2000, Physical Review Letters, 84, 3760

\bibitem[\protect\citeauthoryear{{Spitzer}}{{Spitzer}}{1987}]{Spitzer}
{Spitzer} L.,  1987, {Dynamical evolution of globular clusters}.
Princeton, NJ, Princeton University Press

\bibitem[\protect\citeauthoryear{{Stoehr}, {White}, {Tormen} \&
  {Springel}}{{Stoehr} et~al.}{2002}]{Stoehr2002}
{Stoehr} F.,  {White} S.~D.~M.,  {Tormen} G.,    {Springel} V.,  2002, \mnras,
  335, L84

\bibitem[\protect\citeauthoryear{Suto}{Suto}{2002}]{Suto}
Suto Y.,  2002, in S. B.,  Hwang C.-Y.,  eds, in "Matter and Energy in Clusters
  of Galaxies"

\bibitem[\protect\citeauthoryear{{Swaters}, {Madore}, {van den Bosch} \&
  {Balcells}}{{Swaters} et~al.}{2003}]{Swaters2003}
{Swaters} R.~A.,  {Madore} B.~F.,  {van den Bosch} F.~C.,    {Balcells} M.,
  2003, \apj, 583, 732

\bibitem[\protect\citeauthoryear{{Tremaine}, {Ostriker} \&
  {Spitzer}}{{Tremaine} et~al.}{1975}]{tos75}
{Tremaine} S.~D.,  {Ostriker} J.~P.,    {Spitzer} L.,  1975, \apj, 196, 407

\bibitem[\protect\citeauthoryear{{Verolme}, {Cappellari}, {Copin}, {van der
  Marel}, {Bacon}, {Bureau}, {Davies}, {Miller} \& {de Zeeuw}}{{Verolme}
  et~al.}{2002}]{Verolme02}
{Verolme} E.~K.,  {Cappellari} M.,  {Copin} Y.,  {van der Marel} R.~P.,
  {Bacon} R.,  {Bureau} M.,  {Davies} R.~L.,  {Miller} B.~M.,    {de Zeeuw}
  P.~T.,  2002, \mnras, 335, 517

\bibitem[\protect\citeauthoryear{{Wilkinson}, {Kleyna}, {Evans} \&
  {Gilmore}}{{Wilkinson} et~al.}{2002}]{Mark2002}
{Wilkinson} M.~I.,  {Kleyna} J.,  {Evans} N.~W.,    {Gilmore} G.,  2002,
  \mnras, 330, 778

\bibitem[\protect\citeauthoryear{{Wilkinson}, {Kleyna}, {Evans}, {Gilmore},
  {Irwin} \& {Grebel}}{{Wilkinson} et~al.}{2004}]{Mark2004}
{Wilkinson} M.~I.,  {Kleyna} J.~T.,  {Evans} N.~W.,  {Gilmore} G.~F.,  {Irwin}
  M.~J.,    {Grebel} E.~K.,  2004, \apjl, 611, L21

\bibitem[\protect\citeauthoryear{{Yoo}, {Chanam{\' e}} \& {Gould}}{{Yoo}
  et~al.}{2004}]{ycg04}
{Yoo} J.,  {Chanam{\' e}} J.,    {Gould} A.,  2004, \apj, 601, 311

\bibitem[\protect\citeauthoryear{{Zhao}}{{Zhao}}{1996}]{Zhao1996}
{Zhao} H.,  1996, \mnras, 278, 488

\end{thebibliography}


\end{document}